%% file: gam_interpretability_v2.0.tex
\documentclass[11pt, a4paper]{article} 

\usepackage[utf8]{inputenc}
\usepackage[english]{babel}
\usepackage{comment}
\usepackage[left=2cm, right=2cm, top=2cm, bottom=2cm]{geometry}
\usepackage{graphicx, subcaption}
\usepackage[width=\textwidth]{caption}
\usepackage{setspace}

\usepackage{authblk}

\usepackage[colorlinks=true,allcolors=blue]{hyperref}

\usepackage{amsmath,amsthm,amssymb,mathabx}
\usepackage{bbm,bm}
\usepackage{mathrsfs}%

\usepackage[autostyle]{csquotes}
\usepackage[
    backend=bibtex,
    style=authoryear,
    natbib=true,
    sortlocale=en\_US,
    url=false,
    doi=true,
    eprint=true,
    isbn=false,
    firstinits=true,
    useprefix=true,
    maxbibnames=10,
    maxcitenames=2,
    dashed=false,
	date = year
]{biblatex}
\renewbibmacro{in:}{}

\AtEveryBibitem{%
  \clearfield{note}%
  \clearlist{language}%
}
\DeclareFieldFormat*{title}{#1}
\DeclareNameAlias{sortname}{last-first}
\DeclareNameAlias{default}{last-first}
\usepackage{enumerate}
\usepackage{url, nth}

\usepackage{algorithm}%
\usepackage{algorithmicx}%
\usepackage{algpseudocode}%
\usepackage{listings}%

\usepackage{multirow}
\usepackage{booktabs,colortbl}
\usepackage[table, dvipsnames]{xcolor}%
\usepackage{diagbox}%
\usepackage{dsfont}

\usepackage[title]{appendix}%

\definecolor{LightCyan}{rgb}{0.88,1,1}
\definecolor{LightBlue}{rgb}{0.96,0.98,1}
\definecolor{LightRed}{rgb}{1,0.74,0.79}
\definecolor{LightGreen}{rgb}{0.88,1,0.88}
\definecolor{LightYellow}{rgb}{1,0.98,0.73}
\definecolor{LightPink}{rgb}{1.0,0.88,0.88}

\newcommand{\R}{\mathbb{R}}

\newcommand{\N}{\mathbb{N}}

\newcommand{\esp}[1]{\mathbb{E}\left[#1\right]}

\newcommand{\Xvec}{\boldsymbol{X}}

\newcommand{\xvec}{\boldsymbol{x}}

\newcommand{\zvec}{\boldsymbol{z}}

\newcommand{\betavec}{\boldsymbol{\beta}}

\newcommand{\systL}{\left\{\begin{array}{ll}}
\newcommand{\systR}{\end{array}\right.}
\newcommand{\matL}{\left(\begin{matrix}}
\newcommand{\matR}{\end{matrix}\right)}
\newcommand{\detL}{\left|\begin{matrix}}
\newcommand{\detR}{\end{matrix}\right|}

\newtheorem{rk}{Remark}[section]

\makeatletter
\renewcommand{\@biblabel}[1]{\ }

\makeatother

\title{Explainable Boosting Machine for Predicting Claim Severity and Frequency in Car Insurance}

\author[1]{Markéta Kr\'upov\'a\footnote{Email: \href{mailto:marketa.krupova@addactis.com}{marketa.krupova@addactis.com}.}}
\author[1]{Nabil Rachdi\footnote{Email: \href{mailto:nabil.rachdi@addactis.com}{nabil.rachdi@addactis.com}.}}
\author[2]{Quentin Guibert\footnote{Email: \href{mailto:guibert@ceremade.dauphine.fr}{guibert@ceremade.dauphine.fr}.}}

\affil[1]{ADDACTIS France, 46 bis Chemin du Vieux Moulin, 69160 Tassin-la-Demi-Lune, France}
\affil[2]{CEREMADE, Université Paris-Dauphine, PSL University, CNRS, 75016 Paris, France}

\begin{document}

\maketitle

\begin{abstract}
With the rapid development of machine learning and deep learning techniques, actuaries and the broader insurance industry face a persistent trade-off between predictive accuracy and interpretability. This paper provides a comprehensive applied assessment of Explainable Boosting Machines (EBM) in a car insurance framework, focusing on claim frequency and severity modeling. EBM combines the additive structure of generalized additive models (GAM) with a cyclic gradient boosting algorithm, resulting in a glass-box model whose predictions are interpretable by design.
Using real-world data, we empirically illustrate its practical relevance and compare EBM with modern benchmark models used in non-life insurance pricing. The evaluation considers (i) out-of-sample predictive accuracy, including Murphy diagrams and Bregman dominance tests, and (ii) calibration assessment using T-reliability diagrams and Murphy's score decomposition. Finally, we highlight the link between EBM predictions and Shapley values, showing how predictions can be transparently decomposed into exact main and pairwise interaction effects, providing actionable insights beyond predictive performance.
\end{abstract}

\section{Introduction}\label{sec1}
Recent years have been marked by a rapid expansion of artificial intelligence (AI), machine learning (ML), and deep learning (DL) methods in the insurance industry. Actuaries are naturally affected by this evolution, as these tools are increasingly integrated into pricing and risk modeling practices \citep{eling_impact_2022, richman_ai_2024}. 

While traditional generalized linear models (GLMs) \citep{nelder_generalized_1972, mccullaghnelder89} have long been the standard actuarial framework in non-life insurance, more flexible approaches such as gradient boosting machine (GBM) methods, random forests and neural networks have demonstrated superior predictive performance by capturing complex nonlinear relationships \citep{henckaerts_boosting_2021, poufinas_machine_2023, power_flexible_2024, holvoet_neural_2023}. These modern techniques have enriched insurers’ methodological toolbox, enabling the development of competitive pricing models that more accurately reflect underwritten risks, particularly in high-dimensional data.
Despite their improved performance, many ML and DL models can have the disadvantage of being black-boxes,  making it difficult to interpret and provide transparency in their results, see e.g. \citet{pasquale_black_2015}. This is a major challenge in insurance, given the regulatory obligations related to personal data protection (GDPR) \citep{ocde_impact_2020} and individualized risk analysis. Like other highly regulated industries, insurance must ensure the explainability, fairness, and transparency of algorithms \citep{owens_explainable_2022, charpentier_insurance_2024}. Insurers must be able to justify premium differences and ensure that pricing models do not introduce unintended discrimination \citep{lindholm_discrimination-free_2022, frees_discriminating_2023}. Consequently, fully interpretable models such as GLMs remain widely implemented in practice, despite the superior predictive performance of black-box models with high-dimensional feature effects.

To mitigate this trade-off between performance and interpretability, the literature has developed various explainable artificial intelligence (XAI) tools \citep{molnar_interpretable_2020}. These include intrinsically interpretable models, such as linear or additive models, as well as post-hoc explanation techniques such as Accumulated Local Effects (ALE), Local Interpretable Model-agnostic Explanations (LIME), and SHapley Additive exPlanations (SHAP) \citep{lorentzen_peeking_2020, delcaillau_model_2022}. Another approach consists of approximating complex predictive models using interpretable surrogate models \citep{henckaerts_when_2022}. While these approaches enhance transparency, post-hoc explanations and surrogate approaches may require additional modeling steps and do not always guarantee full faithfulness to the original predictive model. In this context, models that are interpretable by construction while retaining competitive predictive performance are of particular interest for actuarial applications.

In this paper, we provide an applied assessment of Explainable Boosting Machines (EBM) in a non-life insurance pricing context, focusing on claim frequency and severity. Originally introduced by \citet{lou_intelligible_2012, caruana_intelligible_2015} and further implemented by \citet{nori_interpretml_2019}, EBM builds upon generalized additive models (GAMs) \citep{hastie_generalized_1990} and combines them with bagged-boosted decision trees to estimate interpretable shape functions. This approach aligns with a recent stream of literature focused on AI models that are both accurate and directly interpretable \citep{rudin_stop_2019,saleem_explaining_2022}. To achieve this, EBM employs a cyclic gradient boosting procedure, inspired by classical cyclic coordinate descent methods \citep{luenberger_linear_2021}, which sequentially updates feature-specific components. Cyclic boosting schemes are also used in distributional modeling frameworks \citep{chevalierPointProbabilisticGradient2025a}, such as GAMboostLSS \citep{mayr_generalized_2012, hofner_gamboostlss_2016} and the multi-parameter GBM (cyclic GBM) of \citet{delong_cyclic_2023}, where distinct parameters of a response distribution are updated sequentially. However, the role of cyclic training in EBM is conceptually different, since the sequential updates are designed to preserve an additive and interpretable structure, rather than to estimate multiple distributional parameters. Additionally, EBM allows direct visualization of feature effects and supports the identification of pairwise interactions via the GA$^2$M selection algorithm \citep{lou_accurate_2013}.

Given their flexibility, GLM and GAM frameworks often serve as foundations for the recent development of interpretable models on tabular data \citep{wood_interpretability_2022}. For instance, \citet{richman_LocalGLMnet_2023} proposed LocalGLMnet, where the response is modeled through a local GLM whose coefficients are learned by a feed-forward neural network. Similarly, \citet{zakrisson_tree-based_2025} introduced a tree-based varying coefficient model relying on the cyclic multi-parameter GBM procedure of \citet{delong_cyclic_2023}, while \citet{maillart_distill_2024} developed a GAM incorporating shape functions distilled from an ensemble of decision trees. EBM is also conceptually related to the Neural Additive Model (NAM) \citep{agarwal_neural_2021}. It shares the same additive structure to ensure feature-level interpretability, while using specialized neural architectures to learn shape functions. Extensions to distributional settings have been proposed, such as the Neural Additive Model for Location, Scale, and Shape \citep{thielmann_neural_2024}, which mirrors the objectives of GAMboostLSS by jointly modeling multiple distributional parameters within an additive framework.

The EBM model considered in this paper examines the practical relevance of glass-box models for analyzing claim frequency and severity in car insurance.  Our first contribution is empirical. We conduct a comprehensive comparison of EBM with standard actuarial and ML benchmarks commonly used in non-life pricing, including GLM, Extreme Gradient Boosting (XGB), GAMboostLSS, cyclic GBM, NAM, and LocalGLMnet. The evaluation relies on modern predictive performance, calibration, and scoring tools, such as Bregman dominance criteria, T-reliability diagrams, and Murphy decomposition scores. This analysis highlights the strengths and limitations of EBM relative to established alternatives and contributes to enriching the toolbox of interpretable predictive models available to actuaries. Despite its practical appeal and ease of implementation, EBM has, to the best of our knowledge, received limited attention in the actuarial pricing literature. Our second contribution concerns interpretability. We empirically investigate the interpretability properties of EBM in frequency and severity modeling. Unlike black-box approaches, EBM is interpretable by construction and does not rely on post-hoc explanation techniques or surrogate models \citep{henckaerts_when_2022}. Its additive structure enables a transparent decomposition of predictions into individual and pairwise feature contributions. In particular, we emphasize the close correspondence between these shape functions and SHAP values.

This paper is organized as follows. Section~\ref{sec:ebm} introduces the EBM model in the context of non-life insurance pricing. Section~\ref{sec:data_method} presents the methodological framework used to evaluate the EBM approach on a real-world car insurance dataset, for both claim frequency and severity. The corresponding numerical results are discussed in Section~\ref{sec:app_num_results}. Finally, Section~\ref{sec_ccl} concludes the paper.

\section{Methodological framework}
\label{sec:ebm}

The EBM model, introduced by \citet{lou_intelligible_2012} and first empirically evaluated on real data by \citet{caruana_intelligible_2015}, can be viewed as a GAM with univariate and bivariate components learned through ML techniques.
In this section, we describe the methodological framework of the EBM model. We first present the insurance pricing framework and introduce the main notations in Section~\ref{subsec:notations}. The components of the EBM model are then detailed in Section~\ref{subsection:ebm-method}. Finally, we discuss its tuning strategy and interpretability properties in Section~\ref{subsec:discuss_ebm}.

\subsection{Pricing framework and notation}
\label{subsec:notations}

In all that follows, we consider a target space $\mathcal{Y}$, e.g. $\mathcal{Y} = \{0,1\}$ for classification or $\mathcal{Y} = \mathbb{R}$ for regression, and a $p$-dimensional feature space $\mathcal{X}$, where $p \in \mathbb{N}^\star$.
Let $Y \in \mathcal{Y}$ denote the target variable and $X_1,...,X_p$ the $p$ features in $\mathcal{X}$. We denote $\mathbf{x} = (x_1, \ldots, x_p)$ an observation of the variable $\mathbf{X} = (X_1, \ldots, X_p)$ and $y$ an observation of the target variable $Y$. 
Our context is one of supervised learning with a statistical sample of $n \in \N^\star$ observations $\mathcal{D} = \left\{(\Xvec_i,Y_i) \, ; \, i \in  I \right\}$ of an unknown joint law $\mathcal{P}$ on $\mathcal{X} \times \mathcal{Y}$, where $\mathcal{X} \subset \mathbb{R}^p$, $\mathcal{Y} \subset \mathbb{R}_{+}$ and $I=\{1, \ldots, n\}$. The chosen framework is therefore that of regression: for the claim severity, positive real values is predicted, while for the claim frequency, natural integers is targeted. For clarity, bold notation is used for vectors and matrices.

A prediction rule is a measurable function $F: \mathcal{X} \rightarrow \mathcal{Y}$ which associates the output $F(\mathbf{x})$ with the input $\mathbf{x} \in \mathcal{X}$. 
To measure the quality of the prediction, a non-negative consistent loss function $L : \mathcal{Y} \times \mathcal{Y} \rightarrow \mathbb{R_+}$ such that $L(y,y) = 0$ and $L(y,y') > 0$ for $y \neq y'$ is introduced \citep{gneitingMakingEvaluatingPoint2011}. Given a loss function $L$ and a prediction rule $F$, the generalization risk or error is defined by $R_{\mathcal{P}}(F) = \mathbb{E}_{(\mathbf{X},Y) \sim \mathcal{P}}[L(Y,F(\mathbf{X}))]$ and describes the average behavior of the loss function. 

In non-life insurance applications, the response variable is commonly assumed to belong to the exponential dispersion family (EDF), which encompasses most standard distributions used in practice, including the Gaussian, binomial, Poisson, and gamma distributions \citep{ohlsson_non-life_2010, wuthrich_statistical_2023}. In this setting, the unit deviance naturally arises as a consistent loss function for the mean functional. In the numerical applications presented in Section~\ref{sec:app_num_results}, we therefore specify the loss functions minimized during model training using the deviance of the corresponding GLM formulations for claim frequency and claim severity. Formally, the loss function calculated on a generic dataset $\mathcal{D}$ of size $n$ is expressed as follows
\begin{equation} \label{eq:deviance}
L(\mathbf{y}, F(\mathbf{x})) = D(\mathbf{y}, F(\mathbf{x})) = - 2 \ln{\left( \frac{\mathcal{L}(F(\mathbf{x}))}
{{\mathcal{L}(\mathbf{y})}}\right)},
\end{equation}
where $\mathcal{L}$ denotes the likelihood function.

\subsection{Explainable Boosting Machine}
\label{subsection:ebm-method}

Although rarely explored in empirical actuarial research, the smooth functions in a GAM can be replaced by more general shape functions, such as regression trees or decision tree ensembles. In this section, we briefly review the approach introduced by \citet{lou_intelligible_2012}, which enables the construction of GAMs with predictive performance comparable to high-capacity models, such as random forests, boosted trees, or neural networks, while retaining intrinsic interpretability.

\subsubsection{Classical GAM formulation}
\label{subsection:gam}

GAMs \citep{hastie_generalized_1990} are an extension of GLMs \citep{nelder_generalized_1972, mccullaghnelder89}, where the expectation of the response variable $Y$ is explained as a function of a linear combination of predictors. With a single observation $y_i$ of the random variable $Y_i$, $i\in I$, a GLM introduces a regression relationship between $\esp{Y_i}$, and $\mathbf{x}_i$, which assumes for all $\mathbf{\beta} \in \R^{p}$ and for all $i \in I$
\begin{equation}
\label{eq:GLM:linkfun}
g(\esp{Y_i \vert \mathbf{X}_i = \mathbf{x}_i}) =
 \beta_0 + \langle \mathbf{x}_i , \mathbf{\beta}  \rangle \;,
\end{equation}
where $\langle.,.\rangle$ denotes the scalar product, an intercept $\beta_0$, and $g$ is typically a bijective link function. 

For a classical one-dimensional GAM, the effect of each predictor is modeled through an individual additive contribution. In this setting, the deterministic component in~Eq.~\eqref{eq:GLM:linkfun} becomes
\begin{equation}
\label{eq:GAM}
g(\esp{Y_i \vert \mathbf{X}_i = \mathbf{x}_i}) = \beta_0 + 
\sum_{j=1}^{p} \, f_j(x_{i,j}) \;,
\end{equation}
where each shape functions $f_j$, $j \in \{1, \ldots, p\}$ captures the marginal effect of predictor $x_{i,j}$ on the response.
The linear relationship expressed in~Eq.~\eqref{eq:GAM} allows for an easy interpretation of the effects of different features in an additive form. In particular, the use of the logarithmic link function in the case of the gamma and Poisson distributions leads to a multiplicative structure traditionally used in actuarial modeling, which corresponds to a solution making the predicted tariff easily interpretable.

\begin{rk}		
	Traditionally, each feature is associated with a potentially complex non-linear effect, while relating a single attribute to the response variable \citep{wood_interpretability_2022}. This unidimensional specification of shape functions is standard in the actuarial literature, where smooth functions are primarily introduced to capture non-linear effects of continuous covariates \citep{denuit_effective_2019}. In non-life insurance pricing, various smoothing spline approaches \citep{wood_generalized_2006} have been widely adopted, see e.g. \citet{denuit_non-life_2004}, or for mortality graduation, see e.g. \citet{denuit_risk_2018}. For implementation details regarding the fitting of GAMs with spline functions, we refer the reader to \citep{wood_mgcv_2023}.
\end{rk}

Similarly to GLM, interaction components can be added to the formula~\eqref{eq:GAM} to capture multivariate relationships. For instance, a bivariate smooth function $f_{(j_1,j_2)}(x_{i,j_1},x_{i,j_2})$ is used to capture spatial effects, see. e.g. \citet{henckaerts_data_2018}.

\subsubsection{GAM representation of the EBM}
\label{subsubsec:gam_shape}

Through the estimation of univariate shape functions using GBM, EBM bridges ML performance and the structural clarity of classical GAMs. It belongs to the class of glass-box models that aim to balance predictive accuracy and interpretability.

Several tree-based specifications can be considered for the shape functions \citep{lou_intelligible_2012}, including single decision trees, e.g. CART  \citep{breiman_classification_1984}, independently trained trees combined through bagging, sequentially trained trees aggregated via boosting, and hybrid bagged-boosted procedures. Compared with spline-based approaches, these tree-based shape functions generally provide greater flexibility and predictive accuracy, while naturally handling non-continuous features. In particular, \citet{lou_intelligible_2012} report strong empirical performance for the bagged-boosted variant, which is the implementation adopted in EBM.

In addition, \citet{lou_accurate_2013} efficiently extend this approach with two-dimensional interaction shape functions called the GA$^2$Ms method, which remains interpretable by plotting heatmaps. 
As a shape function can be applied to all features,
the model in~Eq.~\eqref{eq:GAM} can then be expressed in the general form
\begin{equation}
\label{eq:GAM-2D}
g(\esp{Y_i \vert \mathbf{X}_i = \mathbf{x}_i}) = \beta_0 + 
\sum_{j=1}^{p} \, f_j(x_{i,j})+
\sum_{j_1=1}^{p} \sum_{j_2\neq j_1}^{p} \, f_{(j_1,j_2)}(x_{i,j_1}, x_{i,j_2})
, \quad \forall i \in I.
\end{equation}
Within the GA$^2$M framework, the authors also propose and empirically assess a feature interaction detection algorithm, denoted FAST, aimed at the computationally efficient selection of relevant pairwise interactions. In EBM, this procedure is applied after the estimation of the main effects and is used to select the pairwise interaction terms to be included in the additive predictor.

\begin{rk}
While the inclusion of bivariate interaction terms may improve predictive performance, GAMs with interactions are known to suffer from identifiability issues and non-uniqueness in the decomposition of effects. This limits interpretability, as the marginal contribution of a given variable can change when interaction terms are introduced. Addressing this issue is beyond the scope of this work and we refer for instance to \citet{lengerichPurifyingInteractionEffects2020}, who propose a functional ANOVA decomposition to purify the representation of GA$^2$M models.
\end{rk}

\subsection{Boosting algorithm and interpretability considerations}
\label{subsec:discuss_ebm}

We now discuss the EBM training algorithm and position it relative to other similar methods. By preserving an additive structure, this algorithm ensures a meaningful correspondence between its predictions and Shapley values.

\subsubsection{Fitting shape functions with a cyclic boosting algorithm}
\label{subsubsec:fitting_ebm}

\citet{nori_interpretml_2019} proposed an efficient implementation in \texttt{Python} of this framework, based on bagged and boosted shallow CART trees to estimate the shape functions $f_j$ or $f_{(j_1,j_2)}$ in~Eq.~\eqref{eq:GAM-2D}. By construction, the EBM algorithm yields accurate local predictions and is particularly well suited to capture the effects of rare observations.

The training procedure for EBM is based on a cyclic boosting algorithm, similar to that proposed by \citet{wick_cyclic_2019}.
The cyclic boosting algorithm is an iterative procedure that relies on a cyclic coordinate descent algorithm \citep{wright_coordinate_2015,luenberger_linear_2021}, in which each shape function associated with a feature $X_j, j = 1, \ldots, p$, is fitted at one time in a round-robin fashion using a low learning rate. Contrary to usual boosting algorithms \citep{breiman_random_2001} where all features are considered simultaneously, this procedure ensures that the order in which the features are estimated does not matter and limits the effects of collinearity. 

\paragraph{Binning procedure}
The cyclic boosting algorithm involves segmenting each feature into $K$-bins. Continuous features are discretized using either a quantile-based or equi-distribution approach to ensure bins of equal size. For categorical variables, the bins correspond to their distinct categories. For bivariate variables resulting from the interaction of two features, the bins represent the interactions of the bins of the two combined variables. Let $b_{j,k}, \; k  \in \{1, \ldots, K\}$ denote the $k$-th bin associated with the $j$-th feature $X_j$. For each bin, we calculate its weight $w_{j,k}$ as the number of observations included in this bin. The number $K$ of bins is identical for each univariate feature. Note that the bins are not updated during the training procedure.

\paragraph{Cyclic boosting algorithm for EBM}
Hereafter, we outline the main steps of the EBM algorithm. To simplify the notation and without loss of generality, we omit the bivariate interaction terms in~Eq.~\eqref{eq:GAM-2D}. Indeed, the binning and fitting procedure described here applies in the same way to univariate and bivariate shape functions.
Using the notation introduced in Section~\ref{subsec:notations}, the prediction rule can be written as
\begin{equation}\label{eq:predict_gam}
F(\mathbf{x}_i) = g^{-1}\left( \mathfrak{F}(\mathbf{x}_i) \right)
= g^{-1}\left(\beta_0 + \sum_{j=1}^{p}{f_j(x_{i,j})} \right),
\end{equation}
where $\mathfrak{F}(\mathbf{x}_i)$  denotes the score of observation $i \in I$. When bivariate interaction terms are considered, the EBM procedure first fits a model including only univariate shape functions. Significant pairs of interacting features are then selected using the FAST algorithm proposed by \citet{lou_accurate_2013}. In a second step, the model is retrained by incorporating the selected interaction components.

\begin{rk}
As a consequence of the use of regression trees and a binning procedure in the GA$^2$M framework, both univariate and bivariate shape functions take the form of step functions. This representation enhances interpretability, as the contribution of each variable is explicit and easily readable.
\end{rk}
Let $t \in \{1, \ldots, T\}$ denote the current iteration and initially set each component of the shape functions to zero. Algorithm~\ref{algo:cyclic_boosting} is initialized by determining the constant that minimizes the loss function. 
At each boosting iteration $t$, each feature $X_{j}$, $j \in \{1, \ldots, p\}$, is partitioned into bins $b_{j,k}, \; k  \in \{1, \ldots, K\}$, allowing shape functions to be trained independently in a cyclical manner. Pseudo-residuals are then computed using the negative gradient of the loss function for all observations $i$ such that $x_{i,j}$ belongs to bin $b_{j,k}$
\begin{equation}\label{eq:pseudo_residual}
\epsilon_i^{(t)} = - \left. {\frac{\partial L(y_i, F(\mathbf{x}_i))}{\partial F(\mathbf{x}_i)}}
\right|_{F(\mathbf{x}_i) = F^{(t-1)}(\mathbf{x}_i)}.
\end{equation}

\begin{algorithm}
\caption{Cyclic boosting algorithm for EBM with regression tree as shape function}
\centering
\footnotesize
\label{algo:cyclic_boosting}
\begin{algorithmic}[1]
\State \textbf{Input:} training data $(\mathbf{x}_i, y_i)_{i=1, \ldots, n}$, $\nu$, $T$, $d$, $L$\; 
\State \textbf{Initialization} 
\State Set $f_{j,k}^{(0)} (\mathbf{x}_i)= 0$,  for $k \in \{1, \ldots, K\}$, $j \in \{1, \ldots, p\}$\; 
\State Set $F^{(0)}(\mathbf{x}_i) = c^\star$, where $c^\star = \arg\min_c \sum_{i=1}^n L(y_i, c)$\;
\State Compute $\beta_0 = g(c)$ \;
\For{$t = 1, \ldots , T$}
	\For{$j = 1, \ldots , p$}
		 \For{$k = 1, \ldots , K$}
		 	\State Compute pseudo-residuals in Eq.~\eqref{eq:pseudo_residual} for all $i$ such that $x_{i,j} \in b_{j,k}$
			\State Fit a regression tree of depth $d$ to pseudo-residuals  $(x_{i,j}, \epsilon_i^{(t)})$
			\State  to obtain regions $R_{j,k,r}^{(t)}$ for $r \in \{1, \ldots, M_{j,k}^{(t)}\}$\;
				\For{$r = 1, \ldots , M_{j,k}^{(t)}$}
					\State Compute prediction
				$$
			\gamma_{j,k,r}^{(t)} = \arg\min_\gamma \sum_{i: x_{i,j} \in R_{j,k,r}^{(t)}\cap b_{k,j}} L\left(y_i, 
			g^{-1}(\mathfrak{F}^{(t-1)}(\mathbf{x}_i) + \gamma) \right).
				$$
				\EndFor
			\State Update bin factor based on Eq.~\eqref{eq:update_bin_factor}
		\EndFor
		\State Update shape function $f_{j}^{(t)}(\mathbf{x}_i) = \sum_{k=1}^{K}{f_{j,k}^{(t)}(\mathbf{x}_i) \mathds{1}_{\{ x_{i,j} \in b_{j,k}\}}}$\;
	\EndFor
	 \State Update prediction $F^{(t)}(\mathbf{x}_i) = g^{-1}\left( \beta_0 + \sum_{j=1}^{p}{f_{j}^{(t)}(\mathbf{x}_i} \right)$\;
\EndFor
\State \textbf{Return} $F(\mathbf{x}_i) = F^{(T)}(\mathbf{x}_i)$\;
\end{algorithmic}
\end{algorithm}

Then, EBM fits a bagged-boosted shallow decision tree of depth $d$ to these pseudo-residuals. This results in a partitioning of the training subset defined by $b_{j,k}$ into $M_{j,k}^{(t)}$ regions based on different features. Each factor $f_{j,k}^{(t)}(\mathbf{x}_i)$ represents the contribution to the score of feature $X_j$ for observations $i$ in bin $b_{j,k}$. These factors are updated in each region $R_{j,k,r}^{(t)}$ by adding a constant $\gamma_{j,k,r}^{(t)}$, whose value is determined by locally minimizing the loss function. The learning rate of the method is controlled by the hyperparameter $\nu$, which helps mitigate the risk of overfitting. The factors $f_{j,k}^{(t)}(\mathbf{x}_i)$ are computed for feature $X_j$ at iteration $t$ as 
\begin{equation}\label{eq:update_bin_factor}
f_{j,k}^{(t)}(\mathbf{x}_i) = f_{j,k}^{(t -1)} (\mathbf{x}_i)+ \nu \sum_{r = 1}^{M_{j,k}^{(t)}}{\gamma_{j,k,r}^{(t)} 
			\mathds{1}_{\{x_{i,j} \in R_{j,k,r}^{(t)} \cap b_{k,j}\}}}.
\end{equation}
The update of the shape function corresponding to feature $X_j$ is then expressed as the sum of the contributions from each bin $b_{j,k}$
$$
f_{j}^{(t)}(\mathbf{x}_i) = \sum_{k=1}^{K}{f_{j,k}^{(t)}(\mathbf{x}_i) \mathds{1}_{\{ x_{i,j} \in b_{j,k}\}}}.
$$
Finally, the prediction function $F$ is updated using Eq.~\eqref{eq:predict_gam} once a full cycle over all variables has been independently completed. 

The EBM fitting procedure follows the same logic of cycling through the features one at a time. The binning procedure, if performed correctly, enables the prediction of rare events in the data. As optimization is performed locally in each bin $b_{j,k}$, atypical observations are treated separately for each factor $X_j$. The handling of rare effects marks a difference between the EBM algorithm and other ML methods, which tend to over-regularize them.

\begin{rk}
The cyclic estimation procedure implemented in the EBM shares similarities with other GBM methods.
\begin{enumerate}
	\item
The cyclic boosting approach implemented in EBM differs from the component-wise gradient boosting method used in algorithms such as GAMboost \citep{hofnerModelbasedBoostingHandson2014}, which is also designed to be interpretable. Within the framework of GAMs, this method involves specifying base-learners for each variable (e.g., linear terms, B-splines, shallow regression trees) and then applying a standard GBM procedure for training \citep{hothornUnbiasedRecursivePartitioning2006, hothornModelbasedBoosting202010}. Unlike the EBM approach, the gradient descent algorithm in component-wise gradient boosting selects the best-performing base-learner at each iteration, rather than systematically cycling through all variables in a round-robin manner. Consequently, the resulting GAM becomes more challenging to interpret, as it is constructed as an additive combination of numerous trees that may interact in complex ways. Moreover, the lack of explicit control over the partitioning process further complicates interpretability. In contrast, EBMs inherently treat each variable (or pairwise interaction) in isolation, ensuring that the model remains interpretable by design.

	\item
Cyclic boosting procedures are also employed in distributional prediction models, see \citet{chevalierPointProbabilisticGradient2025a} for a detailed discussion of the distinction between pointwise and probabilistic gradient boosting. In this setting, the different parameters characterizing a response distribution are updated sequentially within a cyclic training scheme. The GAMboostLSS model \citep{mayr_generalized_2012, hofner_gamboostlss_2016} jointly models several distributional parameters through a GAM and relies on a cyclic multi-parameter GBM procedure. Each parameter is estimated using appropriate base-learners applied to univariate, and in some cases bivariate (e.g., spatial), covariates, following the same component-wise boosting principle as in GAMBoost. Although both GAMboostLSS and EBM rely on cyclic updating schemes, they serve distinct purposes and are not nested within one another. In EBM, cyclic training is used to estimate separate shape functions for each feature, whereas GAMboostLSS applies component-wise gradient boosting within each boosting cycle to update the parameters of a distributional model. The cyclic GBM proposed by \citet{delong_cyclic_2023} is a multi-parameter GBM closely related to GAMboostLSS. It provides a more general framework in the sense that it allows for the simultaneous updating of arbitrary model parameters that are not necessarily tied to distributional moments. Unlike GAMboostLSS, it does not rely on a GAM structure and does not explicitly target interpretability, in contrast to the EBM framework. Moreover, neither GAMboostLSS nor cyclic GBM incorporates a binning step in their training procedures. We also note that EBM focuses on point prediction through the conditional expectation of the response, which remains the primary functional of interest in many actuarial pricing applications.

	\item
Building on this cyclic GBM approach, \citet{zakrisson_tree-based_2025} introduced a tree-based varying coefficient model (VCM) in which coefficients are estimated locally. In this setting, the multi-parameter GBM is used to train local coefficients of a GLM, denoted $\betavec(\zvec)$ with $\zvec \in \mathcal{Z}$. Equation~\eqref{eq:GLM:linkfun} then becomes
$$
g(\esp{Y_i \vert \mathbf{X}_i = \mathbf{x}_i}) =
 \beta_0 + \betavec(\zvec)^\top \xvec_i.
$$
This approach is closely related to EBM and offers a comparable level of interpretability, including effect estimation and variable selection. In particular, the shape functions $f_j(x_{i,j})$ in EBM play a role analogous to the term $\beta_j(\zvec_i) x_{i,j}$. The vector $\zvec$ acts as an effect modifier and can be freely specified, potentially as a subset or transformation of the explanatory variables $\Xvec$, though it is not required to coincide with them. This flexibility makes it possible to incorporate interaction effects within the model. Compared with EBM, the VCM does not include a FAST-type mechanism for selecting bivariate effects. Moreover, the current \texttt{Python} implementation of the model is not optimized for computational efficiency, which can make the search for significant interaction effects prohibitive on real-world datasets.
Finally, the additional flexibility introduced by the effect modifier $\zvec$ requires careful specification of identifiability constraints, which is not an issue in standard EBM models without interaction effects.
\end{enumerate}
\end{rk}

As discussed above, Algorithm~\ref{algo:cyclic_boosting} relies on a multi-parameter GBM procedure, as proposed by \citet{delong_cyclic_2023} and \citet{zakrisson_tree-based_2025}. When convex loss functions are considered, such as the log-likelihoods from the exponential dispersion family introduced in Section~\ref{subsec:notations}, Proposition~1 of \citet{delong_cyclic_2023} ensures in-sample convergence of the EBM algorithm, provided that the shape functions are learned without bagging. To the best of our knowledge, no theoretical guarantees currently exist for bagged-boosted trees beyond empirical evidence, such as that reported by \citet{lou_intelligible_2012} and \citet{caruana_intelligible_2015}. Moreover, the results of \citet{delong_cyclic_2023} do not establish uniqueness of the limit, nor do they guarantee convergence on a test sample. These important issues are not further developed here and would require a dedicated theoretical investigation.

\begin{rk}\label{rk:early_stopping_criterion}
As with other GBM methods, the number of boosting iterations in EBM is controlled through an early stopping criterion \citep{zhangBoostingEarlyStopping2005}. This parameter stops the training process after a predefined number of consecutive iterations without sufficient improvement in the loss function, according to a specified tolerance threshold. In the current EBM implementation provided by the \textbf{InterpretML} package \citep{nori_interpretml_2019}, early stopping is defined globally, meaning that it applies jointly to all shape functions. This contrasts with the approaches of \citet{delong_cyclic_2023,zakrisson_tree-based_2025}, who propose dimension-wise early stopping criteria. Consequently, the global early stopping parameter in EBM involves a trade-off between limiting overfitting and providing enough iterations for the model to empirically assess whether shape functions are effectively null.
Extending the EBM framework to incorporate component-wise early stopping, defined separately for each explanatory variable $X_j$, would be a natural and feasible improvement. In this case, Algorithm~3 of \citet{delong_cyclic_2023} could be directly leveraged.
\end{rk}

\subsubsection{Relationship between EBM shape functions and Shapley values}
\label{subsubsec:relation_shapley}

As discussed above, GAMs form a class of models that are naturally interpretable. In this section, we highlight an appealing property of GAMs that connects them to Shapley values, a widely used class of post-hoc explanation methods. More specifically, \citet{bordtShapleyValuesGeneralized2023} introduce the $n$--Shapley values, which extend both the SHAP feature attributions \citep{lundbergUnifiedApproachInterpreting2017b} and the Shapley Interaction Values \citep{lundbergLocalExplanationsGlobal2020} to interactions of order up to $n$.

Let $\mathcal{C} \subseteq \mathcal{Q}$, where $\mathcal{Q} = \{1, \ldots, p \}$ denotes the set of $p$ input variables, and let $v_{\Xvec} : 2^{\mathcal{Q}} \rightarrow \mathbb{R}$
be a given value function satisfying the classical Shapley axioms (efficiency, symmetry, the dummy player property, and linearity), see e.g. \citet{wuthrichAIToolsActuaries2026}[Chapter~9]. The $n$-Shapley Values $\Phi_{\mathcal{C}}^n$ introduced by \citet{bordtShapleyValuesGeneralized2023} are defined for coalitions $\mathcal{C}$ of cardinality at most $n$ and attribute contributions to groups of up to $n$ variables. For an observation $\xvec \in \mathcal{X}$, they are defined as follows
\begin{equation}
\label{eq:n_shapley_definition}
\Phi_{\mathcal{C}}^n(\xvec)=\sum_{k=0}^{n-|\mathcal{C}|}\sum_{K\subset \mathcal{Q} \setminus \mathcal{C},\,|K|=k} B_k\,\Delta_{\mathcal{C}\cup K}(\xvec),
\end{equation}
where $B_k$ denotes the $k$-th Bernoulli number\footnote{See for example \url{https://mathworld.wolfram.com/BernoulliNumber.html}.} and $\Delta_{\mathcal{C}}(\xvec)$ is the Shapley Interaction Index \citep{grabischAxiomaticApproachConcept1999}, defined as
\begin{equation}
\label{eq:delta_S}
\Delta_{\mathcal{C}}(\xvec) =
\sum_{\mathcal T\subset \mathcal{Q} \setminus \mathcal{C}}
\frac{(p-|\mathcal T|-|\mathcal{C}|)!|\mathcal T|!}{(d-|\mathcal{C}|+1)!}
\sum_{\mathcal L\subset \mathcal{C}}(-1)^{|\mathcal{C}|-|\mathcal L|} v_{\Xvec}(\mathcal L\cup \mathcal T).
\end{equation}
The quantity  $\Phi_{\mathcal{C}}^n$  defined in~Eq.~\eqref{eq:n_shapley_definition} recovers classical Shapley-based explanation concepts as special cases. In particular, for $n = 1$, $\Phi_{\mathcal{C}}^{1}$ coincides with the traditional one-dimensional SHAP feature attributions, while for $n = 2$ it corresponds to the Shapley Interaction Index in~Eq.~\eqref{eq:delta_S}.

Writing $\Xvec = \left( \Xvec_{\mathcal{C}}, \Xvec_{\mathcal{Q} \setminus \mathcal{C}} \right)$ to denote the decomposition of the covariate vector into components inside and outside the coalition $\mathcal{C}$, two main choices of value functions are commonly considered in the literature:
\begin{enumerate}
	\item the observational SHAP value function,
$$
v_{\Xvec}(\mathcal{C}) 
= \esp{ F(\Xvec) \mid \Xvec_{\mathcal{C}} = \xvec_{\mathcal{C}} },
$$
	\item the interventional SHAP value function,
\[
v_{\Xvec}(\mathcal{C}) 
= \esp{ F(\Xvec) \mid do(\Xvec_{\mathcal{C}} = \xvec_{\mathcal{C}}) },
\]
where $do(X = x)$ denotes the causal \emph{do-operator} \citep{pearlDoCalculusRevisited2012}.
\end{enumerate}

\citet{bordtShapleyValuesGeneralized2023}[Theorem 8] show that Shapley-based explanations defined in~Eq.~\eqref{eq:n_shapley_definition} are faithful to GAMs of order $n$. As a direct consequence, this result applies in particular to pairwise interaction models of the GA$^2$M type, such as the model defined in~Eq.~\eqref{eq:GAM-2D}. This faithfulness result holds for interventional SHAP value functions, and also for observational SHAP value functions when the individual features are independent random variables.
In particular, for the interventional SHAP value function, which is widely used in practice due to its computational tractability, this result guarantees that the main effect contribution $f_j(x_{i,j})$ exactly coincides with the SHAP feature attribution $\Phi_{\{j\}}^{1}(\xvec_i)$ whenever the predictive model $F$ is a GAM without interaction terms. In other words, the SHAP decomposition of $\mathfrak{F}(\mathbf{x}_i)$ is identical to the additive decomposition implied by this GAM. Moreover, when the model is a GA$^2$M, the pairwise interaction terms $f_{(j_1,j_2)}(x_{i,j_1}, x_{i,j_2})$ are faithfully recovered by the second-order Shapley attributions $\Phi_{\{j_1,j_2\}}^{2}(\xvec_i)$, while the remaining first-order Shapley Value corresponds to the associated main effect functions.

In contrast, when a generic black-box model $F$, such as a neural network or an XGBoost model, is analyzed using the standard Shapley Values, variable interactions may be present up to a potentially high order $n$. In such settings, the resulting Shapley Values do not provide an exact explanation of the individual effect of a given feature, regardless of the approximation method used to compute them. Achieving an exact functional decomposition would require explicitly accounting for higher-order interaction effects through the computation of the corresponding $n$-Shapley Values. However, the evaluation of these higher-order attributions of a black-box model quickly becomes computationally prohibitive as the interaction order increases.

\section{Empirical assessment of claim frequency and severity modeling}
\label{sec:data_method}

The objective of this section is to provide a fair empirical assessment of the EBM methodology in the context of claim frequency and severity modeling for car insurance. To this end, we compare EBM with a set of benchmark models. The dataset is first described in Section~\ref{subsec:data_desc}. We then present the specification and implementation details of the benchmark models in Section~\ref{subsec:appli_benchmark_models}. Finally, the methodological framework used to evaluate predictive performance and their reliability is detailed in Section~\ref{subsec:app_eval_framework}.

\subsection{Data description}
\label{subsec:data_desc}

The modeling work of this article relies on a car insurance dataset used for research and development purposes within ADDACTIS France. The warranty modeled is the full accidental damage cover over the period from 2012 to 2021. Risk is assumed to be homogeneous over this period. However, a time variable is included in the study to capture the non-structural trend in both claim frequency and severity. Data analysis work precedes the exploitation of these datasets and can be consulted in detail in \citet{krupova_construction_2023}. Particular attention is paid to the feature selection process, with correlation analysis, factor analysis, and supervised feature selection. Based on this previous study, we select a limited number of features, relevant from both a statistical and a business standpoint for the claim frequency and severity modeling. In addition, for computational efficiency, the empirical analysis is performed on a 10\% random sample of the original claim frequency dataset, carefully constructed to preserve the main distributional characteristics of the full database.

For each policyholder $i \in I$, the response variable corresponds to the claim count in the claim frequency dataset and to the claim amount in the claim severity dataset. In addition, the claim frequency dataset includes an exposure-to-risk variable. Summary statistics, together with a detailed presentation of the explanatory variables, are provided in Appendix~\ref{sec:app_data} for both the claim frequency and claim severity datasets.

\subsection{Model configuration and benchmark models}
\label{subsec:appli_benchmark_models}

In this empirical application, we consider two specifications of the EBM model. The first includes only main effects, whereas the second extends the specification by incorporating pairwise interaction effects selected through the FAST algorithm. 

To evaluate the predictive performance of EBM for both claim frequency and claim severity, we compare it with several benchmark models. As a baseline, we consider the GLM, which remains the standard framework in non-life insurance pricing. To ensure a fair comparison, we construct a surrogate GLM designed to approximate the fitted EBM with no interactions. More specifically, we develop a procedure inspired by the \texttt{maidrr} approach of \citet{henckaerts_when_2022}. However, rather than relying on partial dependence effects, we directly extract feature contributions from the EBM shape functions. This EBM-based procedure enables us to identify and remove irrelevant main effects and segment features into clusters that are transformed into categorical factors for inclusion in the GLM specification. The resulting GLM achieves predictive performance close to that of the EBM without interaction terms.

In addition, we consider a range of point prediction models. These include a classical GAM (Section~\ref{subsection:gam}) with cubic regression splines for continuous covariates, the stochastic gradient boosting algorithm XGBoost (XGB) proposed by \citet{chen_2016_xgboost}, the model-based boosting approach GAMboost \citep{hofnerModelbasedBoostingHandson2014}, which relies on component-wise gradient boosting, and the NAM \citep{agarwal_neural_2021, thielmann_neural_2024},  which is structurally close to the EBM framework and learns shape functions through dedicated neural network architectures. We further include the LocalGLMnet model \citep{richman_LocalGLMnet_2023}, a recent interpretable benchmark based on locally estimated GLM coefficients. For all these point prediction models, the loss functions correspond to the Poisson and gamma deviances, see Eq.~\eqref{eq:deviance}, for claim frequency and claim severity, respectively, combined with a logarithmic link function. The use of a log-link in both settings induces a multiplicative tariff structure, thereby preserving interpretability and actuarial relevance.

We also consider two distributional models estimated via cyclic GBM procedures. Specifically, we use GAMboostLSS \citep{mayr_generalized_2012} and the cyclic multi-parameter GBM of \citet{delong_cyclic_2023}. Both models require an explicit distributional assumption. For claim frequency, we assume a negative binomial (NB2) distribution to account for overdispersion in the data and jointly estimate the location and dispersion parameters within a multi-parameter GBM framework. For claim severity, we assume a gamma distribution and sequentially estimate its scale and shape parameters.

Further details on the algorithmic implementation and hyperparameter specification of the EBM models in \texttt{Python} are provided in Appendix~\ref{sec:app_ebm_method}. Implementation details, hyperparameter configurations, and additional methodological information related to the numerical experiments for all competing models are reported in Appendix~\ref{sec:app_model_spec}.

\subsection{Evaluation framework}
\label{subsec:app_eval_framework}

To compare the EBM model with its competitors, we introduce several performance and calibration assessment tools tailored to the insurance pricing framework, both on the training and testing datasets.

\paragraph{Predictive performance.}
To evaluate the quality of the predictive models, we rely on performance measures consistent with our regression setting.
First, for a generic dataset $\mathcal{D}$ of size $n$, we consider the deviance loss defined in~Eq.~\eqref{eq:deviance}, which is a strictly consistent scoring function for the conditional expectation functional \citep{fisslerModelComparisonCalibration2022a}. We also report the Root Mean Squared Error (RMSE) defined as
$$
     \text{RMSE}(\mathbf{y}, F(\mathbf{x})) = \sqrt{\frac{1}{n} \sum_{i=1}^n \, (y_i - F(\mathbf{x_i}))^2}.
$$
To assess whether the observed differences in predictive accuracy are statistically significant, we implement the Diebold--Mariano test \citep{dieboldComparingPredictiveAccuracy2002, holvoet_neural_2023} based on a strictly consistent loss function. More precisely, model $F_B$ is considered statistically more accurate than model $F_A$ if the null hypothesis of the following test is rejected at the $5\%$ significance level
\begin{align*}
H_0 \,&:\, \esp{L(Y, F_A(\Xvec)) - L(Y, F_B(\Xvec))}= 0, \\
H_1 \,&:\, \esp{L(Y, F_A(\Xvec)) - L(Y, F_B(\Xvec))} > 0.
\end{align*}

Second, following \citet{ehm_quantiles_2016,holvoet_neural_2023}, we employ Murphy diagrams to provide a graphical comparison of competing models. These diagrams examine whether the ranking induced by one loss function is robust across a class of consistent scoring functions.
More precisely, for a model $F$ evaluated on the testing dataset $(\mathbf{x}_{i},y_i)$, the Murphy diagram displays the function $(\theta,S_{\theta}(F(\mathbf{x}),y))$,
 where $S_{\theta}(F(\mathbf{x}),y)$ corresponds to the elementary scoring function for a parameter $\theta \in \mathbb{R}$. This estimated quantity is defined as
$$
S_{\theta}(F(\mathbf{x}),y) = \frac{1}{n} \sum_{i = 1}^{n} \, |\theta - y_{i}| \mathds{1}_{\left\{ \min(F(\mathbf{x}_{i}), y_i) \leq \theta < \max(F(\mathbf{x}_{i}),y_i) \right\}}.
$$
According to \citet{ehm_quantiles_2016}, this index compares the predictions of two models. Formally, the predictions from a model $F_{A}$ are said to dominate the predictions of a model $F_{B}$, if and only if, $\forall \, \theta \in \mathbb{R}, \; S_{\theta}(F_{A}(\mathbf{x}),y) \leq S_{\theta}(F_{B}(\mathbf{x}),y)$. The dominance of one model over another, measured through the elementary scoring function, has the advantage of being preserved regardless of the loss function. To formalize the graphical comparison provided by Murphy diagrams, we additionally implement the Bregman dominance test proposed by \citet{denuitComparisonPredictorsPerformance2025}. For two competing models $F_A$ and $F_B$, the test evaluates whether
$
\esp{L(Y, F_A(\Xvec))} \leq \esp{L(Y, F_B(\Xvec))}
$
holds for any Bregman loss function $L$. Formally, model $F_B$ is considered statistically more accurate than model $F_A$ if the null hypothesis of the following test is rejected at the $5\%$ significance level
\begin{align*}
H_0 \,&:\, \mathbb{E}\Big[(t - Y) \big( \mathds{1}_{\{F_B(\mathbf{X}) > t\}} - \mathds{1}_{\{F_A(\mathbf{X}) > t\}} \big) \Big] \le 0, \quad \forall t \in \Omega, \\
H_1 \,&:\, \mathbb{E}\Big[(t - Y) \big( \mathds{1}_{\{F_B(\mathbf{X}) > t\}} - \mathds{1}_{\{F_A(\mathbf{X}) > t\}} \big) \Big] > 0, \quad \text{for some } t \in \Omega,
\end{align*}
where $\Omega = \left[ \min\left( \Omega_A \cup \Omega_B\right), \,  \max\left( \Omega_A \cup \Omega_B\right) \right]$, and $\Omega_A$ and $\Omega_B$ denote the sets of predicted values of $F_A$ and $F_B$, respectively.

\paragraph{Calibration assessment.}
The calibration analysis of actuarial pricing models is an important step for detecting the presence of global or local biases in predictions \citep{fisslerModelComparisonCalibration2022a, gneitingRegressionDiagnosticsMeets2023}. We therefore examine whether our competing models satisfy the property of \emph{auto-calibration}, defined by the condition
\begin{equation}
\label{eq:def_auto_calibration}
F(\Xvec) =  \esp{Y \mid {F(\Xvec)}}.
\end{equation}
This property ensures the absence of cross-subsidization across policies and is a fundamental requirement in insurance pricing. The assessment of auto-calibration is conducted using \emph{T-reliability diagrams} (also referred to as auto-calibration plots), which depict the relationship $s \mapsto \esp{Y \mid {F(\Xvec)} = s}$. 

Constructing such diagrams requires non-parametric regression of observed outcomes on model predictions. 
To this end, we follow the CORP approach proposed by \citet{gneitingRegressionDiagnosticsMeets2023}, which relies on an isotonic regression step applied to the predictions of each model. The resulting T-reliability diagrams can be interpreted as actual-versus-predicted plots after isotonic recalibration, providing a transparent visual diagnostic of auto-calibration property. All calibration diagnostics are implemented using the \texttt{Python} package \textbf{model-diagnostics} \citep{lorentzenchr_model-diagnostics}. This graphical analysis is complemented by a formal statistical test of the auto-calibration property defined in Eq.~\eqref{eq:def_auto_calibration}, as proposed by \citet{denuitTestingAutocalibrationLorenz2024}, which is based on Lorenz and concentration curves 
\footnote{Note that when a model is not auto-calibrated, it is possible to recalibrate the predictions. However, this is not necessarily desirable in a pricing context, as it may lead to the aggregation of several tariff categories \citep{wüthrichIsotonicRecalibrationLow2024a}. At the very least, the balance property can easily be restored by shifting the intercept estimate \citep{lindholmBalancePropertyInsurance2025}. This issue is not further explored in the present article.}.

In addition, we compare our competing models based on the \emph{Murphy's score decomposition} \citep{gneitingRegressionDiagnosticsMeets2023, wuthrichAIToolsActuaries2026}, defined as
$$
\esp{L(Y , {F(\Xvec)})} = UNC(L) - DSC(L) + MSC(L),
$$
where 
$UNC(L) = \esp{L(Y, \esp{Y})}$
denotes the uncertainty component, corresponding to the prediction error of the trivial model that always predicts the global mean,
$DSL(L) = \esp{L(Y, \esp{Y})} - \esp{L(Y, F_{rc}(\Xvec))}$
is the discrimination term, measuring the reduction in uncertainty achieved by the recalibrated predictor $F_{rc}(\Xvec)$, and
$MSC(L) = \esp{L(Y, F_{rc}(\Xvec))} - \esp{L(Y, F(\Xvec))}$
is the miscalibration component, capturing the uncertainty attributable to deviations from the auto-calibration property. All components of the Murphy's score decomposition are computed using the \textbf{model-diagnostics} package.



\section{Numerical results}
\label{sec:app_num_results}

This section presents the numerical results for claim severity. For completeness, the corresponding results for claim frequency are reported in Appendix~\ref{sec:app_results_freq}. Section~\ref{subsec:app_perf} focuses on the predictive performance of the competing severity models. In Section~\ref{subsec:app_reliability}, we examine their auto-calibration properties. Finally, Section~\ref{subsec:interpratabilty} provides an empirical illustration of how EBM predictions can be used to interpret model outputs.
All numerical evaluations are conducted using the \texttt{R} \citep{rsoft25} and \texttt{Python} software environments.

\subsection{Predictive performance for claim severity models}
\label{subsec:app_perf}

We start by comparing the predictive performance of all competing models as defined in Section~\ref{subsec:app_eval_framework}: surrogate GLM (GLMbins), GAM, GAMboost, GAMboostLSS, cyclic multi-parameter GBM (cyc-GBM), EBM without interactions (EBM), EBM with bivariate interactions (EBM$^2$), XGB, NAM, and LocalGLMnet. 

Performance metrics in terms of deviance and RMSE are reported for claim severity in Table~\ref{tab:res_severity}, both in-sample (training sets) and out-of-sample (testing sets).
EBM with interactions slightly outperforms the benchmark models on the testing dataset both in terms of deviance and RMSE, closely followed by cyc-GBM, XGB and LocalGLMnet. We also note that EBM without interactions shows comparable out-of-sample prediction performance to GAMboost and GAMboostLSS. While the reference GLMbins model and GAM achieve similar performance levels, NAM displays lower predictive performance in terms of both deviance and RMSE.

\begin{table}[h!]
\centering
\caption{In-sample and out-of-sample prediction performance in terms of deviance and RMSE for the claim severity models. For each model, the percentage column gives the relative gain (highlighted in \textcolor{teal}{green}) or loss (highlighted in \textcolor{red}{red}) over the reference GLMbins model. \label{tab:res_severity}}
\fontsize{8}{10}\selectfont
\renewcommand{\arraystretch}{1.2} 
\setlength{\tabcolsep}{6pt}
\begin{tabular}{lcccccccccc}
\toprule
\textbf{Model} & \multicolumn{3}{c}{\textbf{Deviance}} & \multicolumn{3}{c}{\textbf{RMSE}} \\
 \cmidrule(lr){2-4} \cmidrule(lr){5-7} 
 & Train & Test & \% & Train & Test & \% \\
\midrule
GLMbins & 0.7478 & 0.7364 & 0 & 2098 & 2011 & 0 \\
GAM & 0.7495 & 0.7365 & \textcolor{orange}{0.0\%} & 2099 & 2012 & \textcolor{orange}{0.0\%} \\
GAMboost & 0.7494 & 0.7353 & \textcolor{teal}{-0.1\%} & 2100 & 2010 & \textcolor{teal}{-0.1\%} \\
GAMboostLSS & 0.7500 & 0.7356 & \textcolor{teal}{-0.1\%} & 2102 & 2010 & \textcolor{orange}{0.0\%} \\ 
cyc-GBM & 0.7408 & 0.7326 & \textcolor{teal}{-0.5\%} & 2087 & 2004 & \textcolor{teal}{-0.3\%} \\
EBM & 0.7491 & 0.7356 & \textcolor{teal}{-0.1\%} & 2100 & 2010 & \textcolor{teal}{-0.1\%} \\
EBM$^2$ & 0.7445 & \textbf{0.7318} & \textcolor{teal}{-0.6\%} & 2091 & \textbf{2004} & \textcolor{teal}{-0.4\%} \\
XGB & 0.7333 & 0.7332 & \textcolor{teal}{-0.4\%} & 2072 & 2006 & \textcolor{teal}{-0.3\%} \\
NAM & 0.7612 & 0.7447 & \textcolor{red}{+1.1\%} & 2115 & 2023 & \textcolor{red}{+0.6\%} \\
LocalGLMnet & 0.7443 & 0.7340 & \textcolor{teal}{-0.3\%} & 2093 & 2008 & \textcolor{teal}{-0.1\%} \\ 
\bottomrule
\end{tabular}
\end{table}

The out-of-sample results of the Diebold--Mariano tests are reported in Table~\ref{tab:DM_severity}. For each row, we test whether the null hypothesis $H_0$ cannot be rejected, i.e., whether model $F_A$ (reported in the row) is not statistically less accurate than model $F_B$. Conversely, a $p$-value below $0.05$ indicates that model $F_B$ (reported in the column) is statistically more accurate than model $F_A$. 

For the claim severity data, these tests do not allow us to conclude that a single model dominates all others in terms of deviance, which is consistent with the minor differences previously observed in RMSE and deviance. In particular, when GLMbins is considered as model $F_A$, the null hypothesis is never rejected, as its specification incorporates insights derived from the EBM structure. When considered as model $F_B$, GLMbins significantly outperforms GAMboostLSS, cyc-GBM, and XGB at the $5\%$ significance level. Furthermore, at the $5\%$ significance level, the results indicate that EBM and EBM$^2$ achieve higher predictive accuracy than GAMboostLSS, cyc-GBM, XGB, and LocalGLMnet. The results of Diebold-Mariano test demonstrate that, despite being primarily designed for interpretability, the EBM and EBM$^2$ models remain highly competitive in terms of predictive performance.

\begin{table}[ht]
\fontsize{8}{10}\selectfont
\centering
\caption{
Results of the Diebold--Mariano tests for the claim severity models on the testing dataset, based on the gamma deviance. For each pair consisting of a row model $F_A$ and a column model $F_B$, the null hypothesis $H_0$ is rejected if the corresponding $p$-value is below $0.05$ (highlighted in \textcolor{teal}{green}). In this case, model $F_B$ is considered statistically more accurate than model $F_A$.
 \label{tab:DM_severity}}
\renewcommand{\arraystretch}{1.3} 
\setlength{\tabcolsep}{6pt}
\begin{tabular} {lcccccccccc}
\toprule
\textbf{Model} & GLMbins & GAM & GAMboost & GAMboostLSS & cyc-GBM & EBM & EBM$^2$ & XGB & NAM & LocalGLMnet \\ \midrule
GLMbins & & 0.668 & 0.757 & 0.971 & 0.963 & 0.212 & 0.452 & 0.977 & 0.857 & 0.944 \\
GAM & 0.332 & & 0.584 & 0.930 & 0.944 & 0.091 & 0.360 & 0.963 & 0.830 & 0.926 \\
GAMboost & 0.243 & 0.416 & & 0.998 & 0.964 & 0.080 & 0.287 & 0.973 & 0.848 & 0.915 \\
GAMboostLSS & \textcolor{teal}{0.029} & 0.070 & \textcolor{teal}{0.002} & & 0.895 & \textcolor{teal}{0.004} & \textcolor{teal}{0.038} & 0.897 & 0.666 & 0.739 \\
cyc-GBM & \textcolor{teal}{0.037} & 0.056 & \textcolor{teal}{0.036} & 0.105 & & \textcolor{teal}{0.026} & \textcolor{teal}{0.029} & 0.470 & 0.210 & 0.277 \\
EBM & 0.788 & 0.909 & 0.920 & 0.996 & 0.974  & & 0.637 & 0.984 & 0.912 & 0.966 \\
EBM$^2$ & 0.548 & 0.640 & 0.713 & 0.962 & 0.971 & 0.363 & & 0.982 & 0.876 & 0.967 \\
XGB & \textcolor{teal}{0.023} & \textcolor{teal}{0.038} & \textcolor{teal}{0.027} & 0.103 & 0.530 & \textcolor{teal}{0.016} & \textcolor{teal}{0.018} & & 0.225 & 0.252 \\
NAM & 0.143 & 0.170 & 0.152 & 0.334 & 0.790 & 0.088 & 0.124 & 0.775 & & 0.569 \\
LocalGLMnet & 0.056 & 0.074 & 0.085 & 0.261 & 0.723 & \textcolor{teal}{0.035} & \textcolor{teal}{0.033} & 0.748 & 0.431 &  \\ \bottomrule
\end{tabular}
\end{table}

\begin{figure}[ht!]
\centering
\begin{subfigure}{.33\textwidth}
  \centering
  \includegraphics[width=\linewidth]{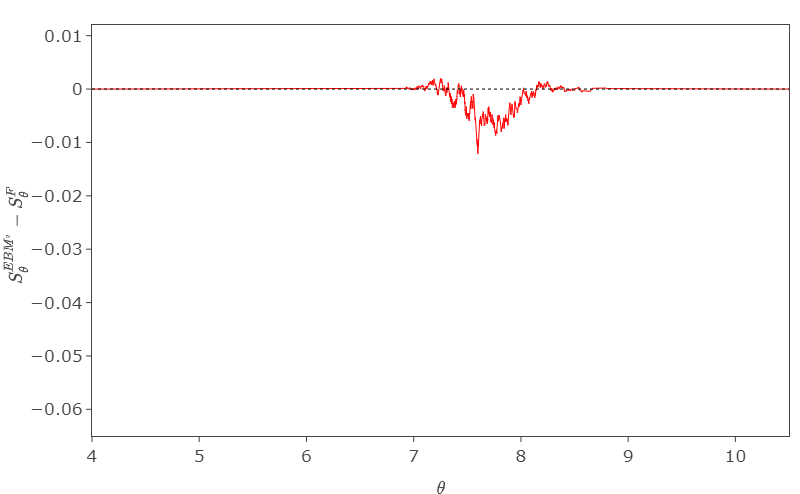}
  \caption{GLMbins}
  \label{fig:murphy_sev_GLM}
\end{subfigure}%
\begin{subfigure}{.33\textwidth}
  \centering
  \includegraphics[width=\linewidth]{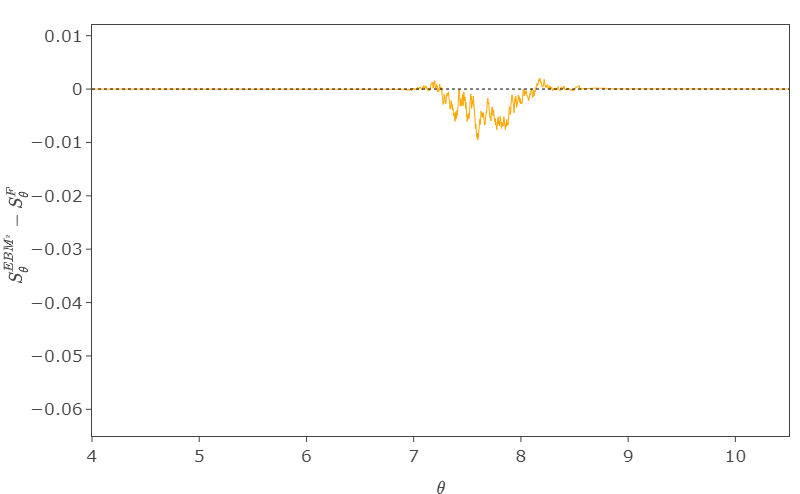}
  \caption{GAM}
  \label{fig:murphy_sev_GAM}
\end{subfigure}%
\begin{subfigure}{.33\textwidth}
  \centering
  \includegraphics[width=\linewidth]{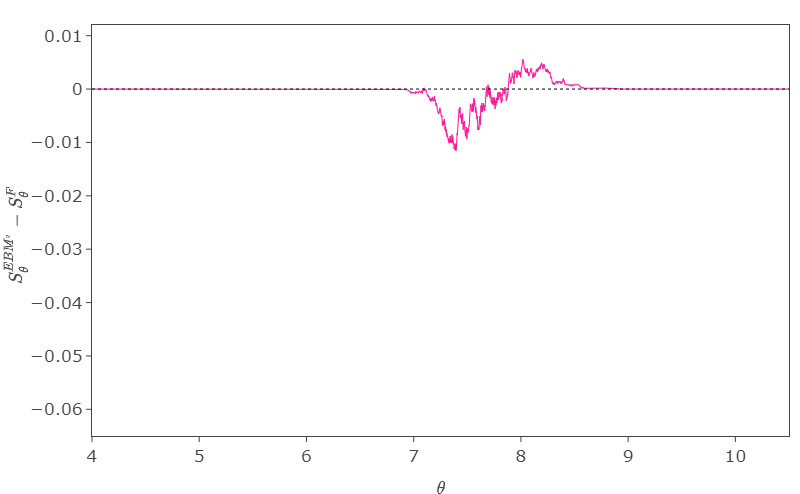}
  \caption{GAMboost}
  \label{fig:murphy_sev_GAM_Boost}
\end{subfigure}
\begin{subfigure}{.33\textwidth}
  \centering
  \includegraphics[width=\linewidth]{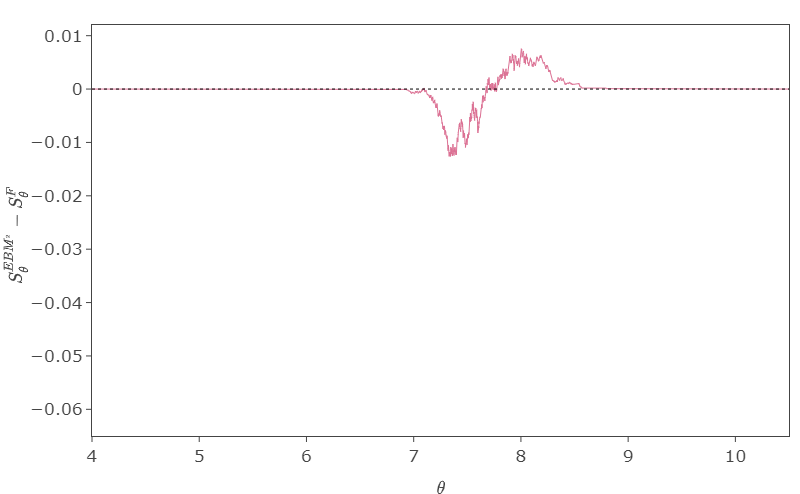}
  \caption{GAMboostLSS}
  \label{fig:murphy_sev_GAM_Boost_LSS}
\end{subfigure}%
\begin{subfigure}{.33\textwidth}
  \centering
  \includegraphics[width=\linewidth]{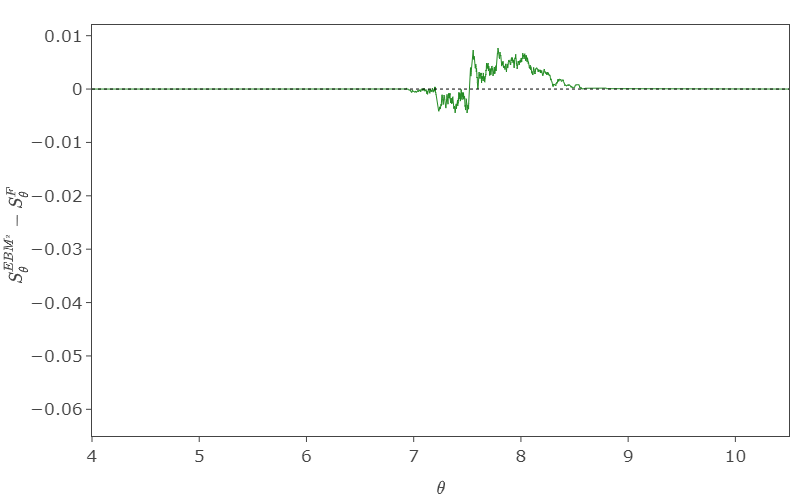}
  \caption{cyc-GBM}
  \label{fig:murphy_sev_GBM}
\end{subfigure}%
\begin{subfigure}{.33\textwidth}
  \centering
  \includegraphics[width=\linewidth]{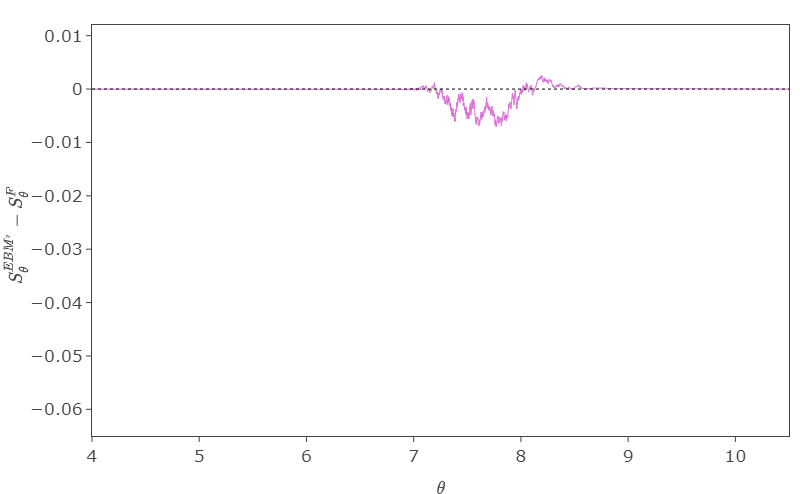}
  \caption{EBM}
  \label{fig:murphy_sev_EBM}
\end{subfigure}
\begin{subfigure}{.33\textwidth}
  \centering
  \includegraphics[width=\linewidth]{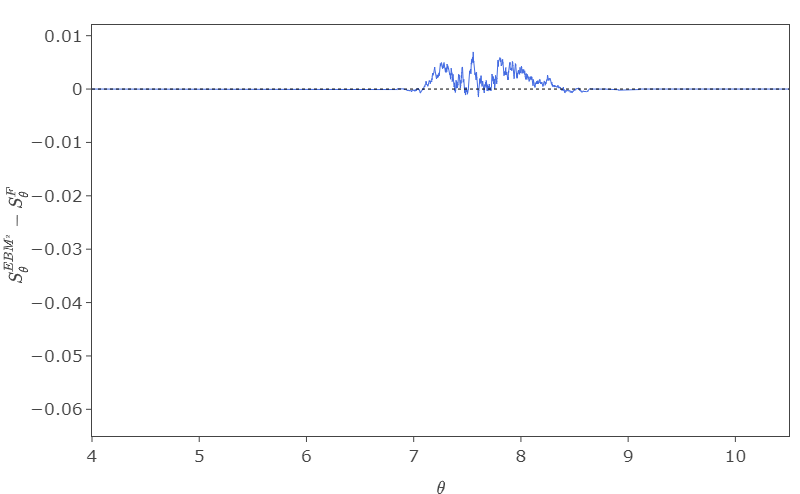}
  \caption{XGB}
  \label{fig:murphy_sev_XGB}
\end{subfigure}%
\begin{subfigure}{.33\textwidth}
  \centering
  \includegraphics[width=\linewidth]{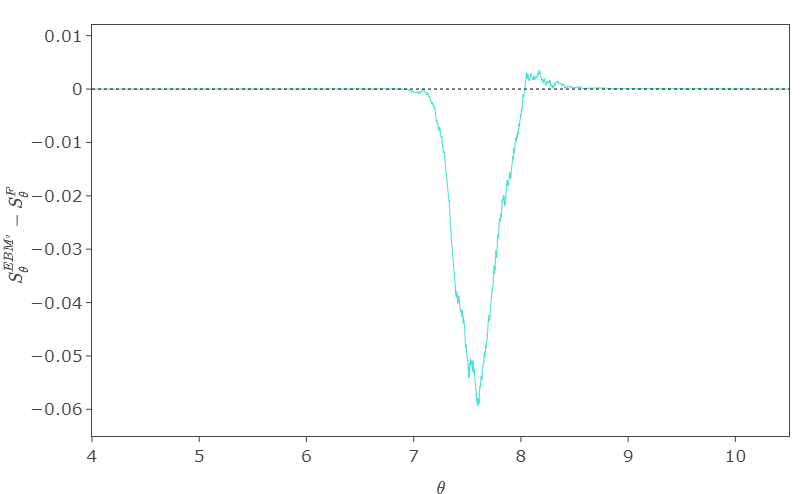}
  \caption{NAM}
  \label{fig:murphy_sev_NAM}
\end{subfigure}
\begin{subfigure}{.33\textwidth}
  \centering
  \includegraphics[width=\linewidth]{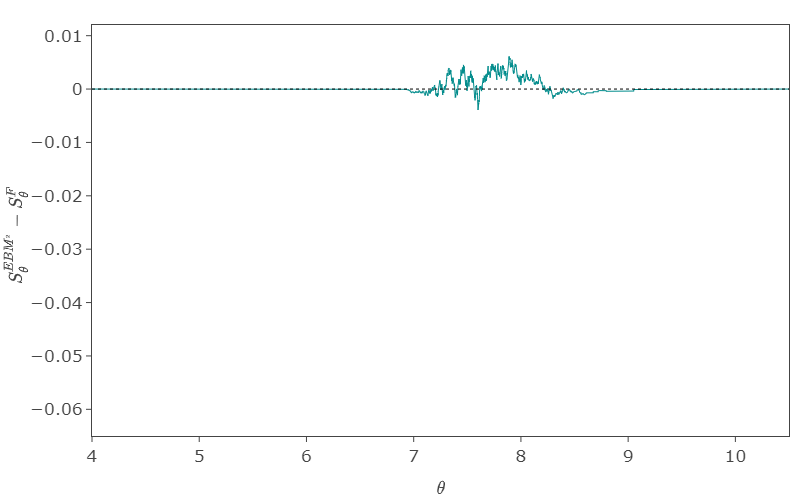}
  \caption{LocalGLMNet}
  \label{fig:murphy_sev_LGLM}
\end{subfigure}
\caption{
Murphy diagrams for the claim severity models on the test dataset,
based on the difference between the elementary scoring function $S_{\theta}$
of the EBM$^2$ model and that of each competing model across the range of
parameter values $\theta$. A positive difference indicates that the competing
model outperforms EBM$^2$, whereas a negative difference indicates that
EBM$^2$ outperforms the competing model.
}
\label{fig:murphy_sev}
\end{figure}

Subsequently, the predictive dominance of EBM$^2$ over the competing
benchmark models is assessed graphically using Murphy diagrams.
Figure~\ref{fig:murphy_sev} presents the difference between the elementary
scoring function $S_{\theta}$ of the EBM$^2$ model and that of each
competing model across the range of parameter values $\theta$, evaluated on
the test dataset. First, in Figure~\ref{fig:murphy_sev_NAM}, the predictions
produced by the NAM model are seen to be almost uniformly dominated by
EBM$^2$. The EBM$^2$ model also appears to dominante GLMbins,
GAM and EBM over a large portion of the range of $\theta$.
The curves associated with GAMboost
(Figure~\ref{fig:murphy_sev_GAM_Boost}) and GAMboostLSS
(Figure~\ref{fig:murphy_sev_GAM_Boost_LSS}) are dominated by EBM$^2$ for
values of $\theta$ approximately between $7.1$ and $7.8$, but dominate it in
the region between $7.9$ and $8.6$. Finally, cyc-GBM
(Figure~\ref{fig:murphy_sev_GBM}), XGB (Figure~\ref{fig:murphy_sev_XGB}) and
LocalGLMNet (Figure~\ref{fig:murphy_sev_LGLM}) appear to dominate EBM$^2$
over a large portion of the range of $\theta$, without achieving uniform
dominance.

Finally, the graphical analysis based on Murphy diagrams is complemented by the out-of-sample Bregman dominance test, reported in Table~\ref{tab:BD_severity}, over a range of predicted values $\Omega$. Similar to the Diebold--Mariano test, this procedure evaluates, for each pair of models $(F_A, F_B)$, whether the null hypothesis $H_0$ cannot be rejected. Accordingly, model $F_B$ (reported in the column) is considered statistically more accurate than model $F_A$ when the corresponding $p$-value falls below $5\%$. The results indicate that no model uniformly dominates all others across the entire range of predictions. Nevertheless, XGB, cyc-GBM, and EBM$^2$ exhibit statistically significant dominance over several competing models.

\begin{table}[ht!]
\fontsize{8}{10}\selectfont
\centering
\caption{
Results of the Bregman dominance tests for the claim severity models on the testing dataset. For each pair consisting of a row model $F_A$ and a column model $F_B$, the null hypothesis $H_0$ is rejected if the corresponding $p$-value is below $0.05$ (highlighted in \textcolor{teal}{green}). In this case, model $F_B$ is considered statistically more accurate than model $F_A$. \label{tab:BD_severity}}
\renewcommand{\arraystretch}{1.3} 
\setlength{\tabcolsep}{6pt}
\begin{tabular} {lcccccccccc}
\toprule
\textbf{Model} & GLMbins & GAM & GAMboost & GAMboostLSS & cyc-GBM & EBM & EBM$^2$ & XGB & NAM & LocalGLMnet \\ \midrule
GLMbins & & 0.518 & 0.344 & 0.590 & \textcolor{teal}{0.018} & 0.488 & \textcolor{teal}{0.004} & \textcolor{teal}{0.030} & 0.998 & 0.190 \\
GAM & 0.456 & & 0.206 & 0.270 & \textcolor{teal}{0.008} & 0.612 & \textcolor{teal}{0.004} & \textcolor{teal}{0.018} & 0.994 & 0.072 \\
GAMboost & 0.756 & 0.978 & & 1.000 & \textcolor{teal}{0.002} & 0.940 & \textcolor{teal}{0.000} & \textcolor{teal}{0.020} & 1.000 & 0.164 \\
GAMboostLSS & 0.940 & 0.936 & 0.528 & & \textcolor{teal}{0.002} & 0.716 & \textcolor{teal}{0.018} & \textcolor{teal}{0.048} & 0.998 & 0.116 \\
cyc-GBM & 1.000 & 0.998 & 0.948 & 0.998 & & 0.984 & 0.224 & 0.860 & 1.000 & 0.658 \\
EBM & 0.932 & 0.300 & 0.494 & 0.848 & \textcolor{teal}{0.006} & & \textcolor{teal}{0.002} & \textcolor{teal}{0.022} & 1.000 & 0.072 \\
EBM$^2$ & 1.000 & 1.000 & 0.800 & 0.964 & 0.780 & 0.986 & & 0.882 & 1.000 & 0.676 \\
XGB & 0.796 & 0.520 & 0.850 & 0.864 & 0.288 & 0.810 & 0.342 & & 1.000 & 0.934 \\
NAM & \textcolor{teal}{0.002} & \textcolor{teal}{0.000} & \textcolor{teal}{0.004} & \textcolor{teal}{0.000} & \textcolor{teal}{0.000} & \textcolor{teal}{0.000} & \textcolor{teal}{0.000} & \textcolor{teal}{0.000} & & \textcolor{teal}{0.000} \\
LocalGLMnet & 0.982 & 0.828 & 0.834 & 0.854 & 0.060 & 0.692 & \textcolor{teal}{0.016} & \textcolor{teal}{0.026} & 0.998 &  \\ \bottomrule
\end{tabular}
\end{table}

\subsection{Calibration assessment}
\label{subsec:app_reliability}

\begin{figure}[ht!]
    \centering
    \includegraphics[width=0.7\linewidth]{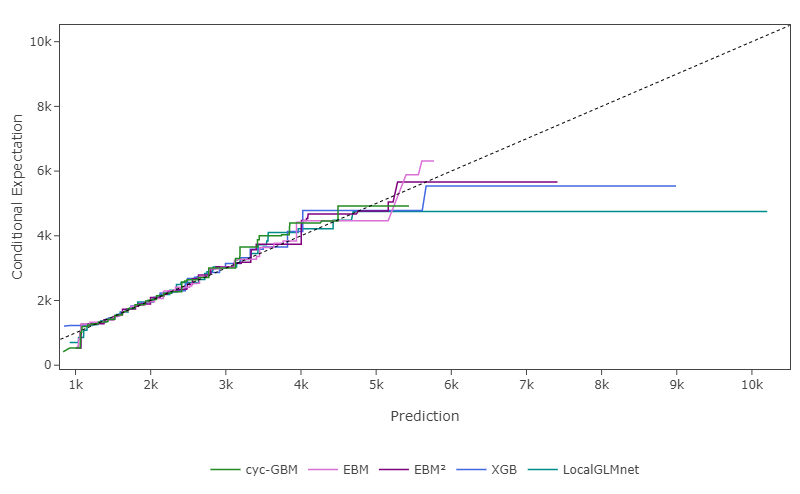}
    \caption{Calibration plot for the claim severity cyc-GBM, EBM, EBM$^2$, XGB and LocalGLMnet models on the testing dataset.}
    \label{fig:isotonic_sev}
\end{figure}

We now examine the auto-calibration properties of the considered models. The out-of-sample calibration assessment based on T-reliability diagrams is presented in Figure~\ref{fig:isotonic_sev} for the five best-performing models in terms of predictive accuracy and dominance, namely cyc-GBM, EBM with and without interactions, XGB, and LocalGLMnet. According to this graphical diagnostic, EBM without interactions and cyc-GBM stand out for their superior calibration quality. However, the latter exhibits some difficulty in accurately reproducing the most extreme values at both ends of the distribution. This graphical analysis is complemented by the auto-calibration test of \citet{denuitTestingAutocalibrationLorenz2024}. Table~\ref{tab:calibration_test_sev} reports the corresponding $p$-values for all competing benchmark models on the testing dataset. The null hypothesis $H_0$ of auto-calibration is not rejected at the $0.5\%$ significance level for all models except NAM, thereby confirming the conclusions drawn from the graphical analysis for the five best-performing models.

\begin{table}[ht!]
\centering
	\caption{Auto-calibration test for the claim severity models on the testing dataset. 
The null hypothesis of auto-calibration is rejected at the $0.5\%$ significance level when $p < 0.005$.}
   \fontsize{8}{10}\selectfont
    \centering
\begin{tabular}{cccc}
\toprule
\textbf{Model} & {\textbf{$p$-value}} & 
\textbf{Model} & {\textbf{$p$-value}} \\
\midrule
GLMbins      & 0.942 & cyc-GBM     & 0.287 \\
GAM          & 0.985 & EBM         & 0.995 \\
GAMboost     & 0.567 & EBM$^2$     & 0.950 \\
GAMboostLSS  & 0.525 & XGB         & 0.422 \\
NAM          & $<0.001$ & LocalGLMnet & 0.204 \\
\bottomrule
\end{tabular}
\label{tab:calibration_test_sev}
\end{table}

Finally, we compare the competing benchmark models based on the Murphy's score decomposition on the testing dataset, as reported in Table~\ref{tab:murphy_dec_sev}. This analysis quantifies how a model discriminates the response variable with respect to the covariates, while separating the contributions of miscalibration, discrimination, and intrinsic uncertainty. The EBM model with interactions exhibits the strongest discrimination ability, whereas the EBM model without interactions appears to be the least affected by miscalibration. When jointly accounting for miscalibration, discrimination, and uncertainty, the EBM$^2$ model achieves the best overall score, followed by cyc-GBM, XGB, and LocalGLMnet.

\begin{table}[!ht]
    \caption{Murphy’s score decomposition for the claim severity models on the testing dataset, based on the deviance loss.}
    \fontsize{8}{10}\selectfont
    \centering
    \renewcommand{\arraystretch}{1.3} 
    \setlength{\tabcolsep}{6pt}
    \begin{tabular} {lccccc}
    \toprule
    \textbf{Model} & \textbf{Miscalibration} & \textbf{Discrimination} & \textbf{Uncertainty} & \textbf{Score} \\ \midrule
    GLMbins & 0.0031 & 0.0696 & 0.8029 & 0.7364 \\ 
    GAM & 0.0029 & 0.0693 & 0.8029 & 0.7365 \\ 
    GAMboost & 0.0029 & 0.0704 & 0.8029 & 0.7353 \\ 
    GAMboostLSS & 0.0032 & 0.0704 & 0.8029 & 0.7356 \\ 
    cyc-GBM & 0.0029 & 0.0731 & 0.8029 & 0.7326 \\ 
    EBM & \textbf{0.0028} & 0.0701 & 0.8029 & 0.7356 \\ 
    EBM$^2$ & 0.0035 & \textbf{0.0746} & 0.8029 & \textbf{0.7318} \\ 
    XGB & 0.0030 & 0.0726 & 0.8029 & 0.7332 \\ 
    NAM & 0.0057 & 0.0638 & 0.8029 & 0.7447 \\ 
    LocalGLMnet & 0.0032 & 0.0720 & 0.8029 & 0.7340 \\ \bottomrule
    \end{tabular}
    \label{tab:murphy_dec_sev}
\end{table}

\subsection{Interpretability analysis}
\label{subsec:interpratabilty}

After assessing the predictive performance of EBM relative to competing models, we now present how this approach provides directly interpretable analyses of its predictions. In the following, we present interpretability tools derived from the model's shape functions for a generic dataset $\mathcal{D}$ of size $n$. In this section, we present only the results for the claim severity model. Equivalent results to those presented in Section~\ref{subsec:tool_interpretability} for the claim frequency models are provided in Appendix~\ref{sec:app_results_freq}.

\subsubsection{Interpretation with shape functions}
\label{subsec:tool_interpretability}

We begin by illustrating interpretability analysis through the EBM shape functions on the training set. 
The main effects and interaction terms provide a transparent description of how individual features influence the predicted outcome. 
Specifically, the overall effect of a feature can be visualized and interpreted via its corresponding shape function $f_j$ (or $f_{(j_1,j_2)}$ for pairwise interactions) of the trained model, as defined in Eq.~\eqref{eq:GAM-2D}.

The global feature importance score, $I_j$, and the corresponding relative feature importance, $RI_j$, of a variable $X_j$ can be computed based on the estimated shape function $f_j$ (or similarly for pairwise interactions $f_{(j_1,j_2)}$) as
\begin{equation}
\label{eq:feat_importance_ebm}
I_j  = \sum_{k=1}^{K}{\frac{w_{j,k}}{\sum_{m=1}^{K}{w_{j,m}}} \displaystyle \left\lvert  \sum_{i=1}^{n}{f_{j,k}(\mathbf{x}_i) \mathds{1}_{\{ x_{i,j} \in b_{j,k}\}}} \right\rvert}, \qquad
RI_j = \frac{I_j}{\sum_{j=1}^{p} I_j},
\end{equation}
where $w_{j,k} = \sum_{i=1}^{n}{\mathds{1}_{\{ x_{i,j} \in b_{j,k}\}}}$ is the weight of bin $b_{j,k}$, i.e. the number of observations in that bin. 
A high value of $I_j$ indicates that the variable $X_j$ contributes strongly to the model's predictions. These indices can also be computed for bivariate interactions and used to rank and select the most relevant pairwise components.

\begin{figure}[h!]
    \centering
    \begin{subfigure}{.33\textwidth}
    	\centering
    	\includegraphics[width=\linewidth]{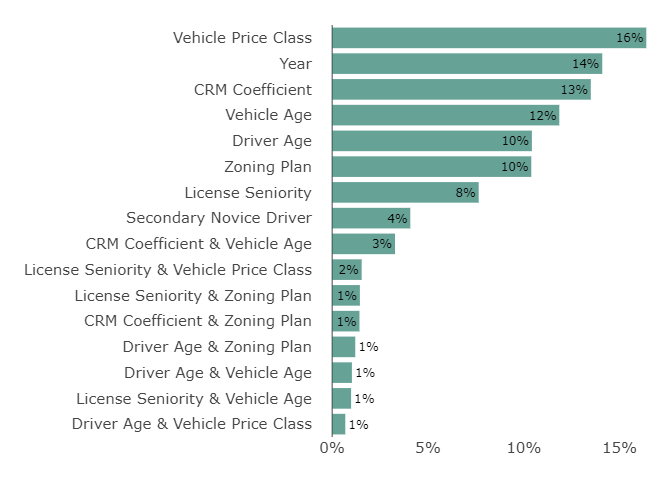}
    	\caption{Relative feature importance $RI_j$}
    	\label{fig:global_EBM_sev}
    \end{subfigure}%
	\begin{subfigure}{.33\textwidth}
  		\centering
  		\includegraphics[width=\linewidth]{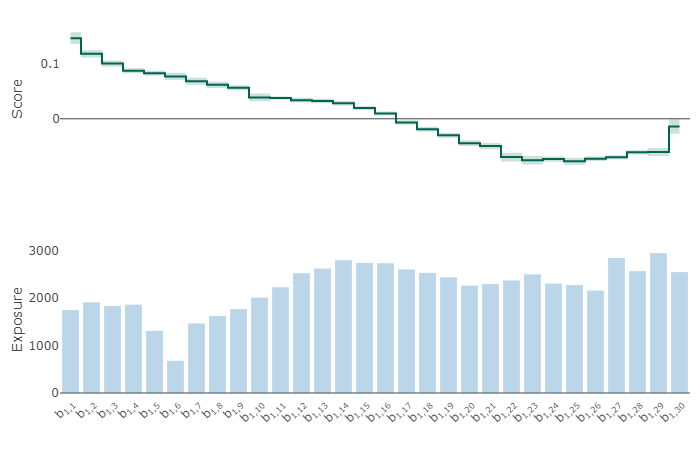}
  		\caption{Driver Age}
  		\label{fig:shape_age_COND_sev}
  	\end{subfigure}%
 	\begin{subfigure}{.33\textwidth}
 		\centering
 		\includegraphics[width=\linewidth]{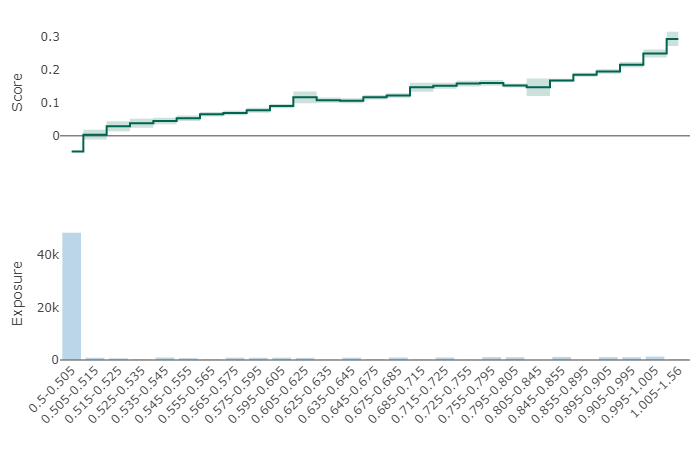}
 		\caption{CRM Coefficient}
 		\label{fig:shape_cof_CRM_sev}
 	\end{subfigure}   
\caption{
Relative feature importance in the EBM$^2$ claim severity model, with features sorted by descending importance (Figure~\ref{fig:global_EBM_sev}), and predicted shape functions $f_j(\mathbf{x}_i)$ for Driver Age (Figure~\ref{fig:shape_age_COND_sev}) and CRM Coefficient (Figure~\ref{fig:shape_cof_CRM_sev}). Shaded areas indicate uncertainty bands estimated via bagging.
}
\label{fig:shape_sev}
\end{figure}

Figure~\ref{fig:shape_sev} presents visualizations derived from the EBM$^2$ shape functions, offering both a global overview of variable importance and a local view of main effects on claim severity. 
Figure~\ref{fig:global_EBM_sev} displays the relative feature importance $RI_j$ for univariate and pairwise interaction terms, previously selected by the FAST algorithm according to their importance. In the claim severity model, Vehicle Price Class is the most influential feature, while CRM Coefficient and Year also contribute substantially. Overall, pairwise interactions have a smaller impact than the main effects of univariate features. This representation nonetheless makes it possible to rank and position the interaction terms. From a practical perspective, such an analysis can be leveraged within a variable selection framework for building a fully interpretable pricing model, such as a GLM. In particular, selected bivariate effects could be reintroduced into the GLMbins specification to potentially enhance its predictive performance. We do not explore this extension further in the present paper.

Figures~\ref{fig:shape_age_COND_sev} and~\ref{fig:shape_cof_CRM_sev} provide detailed visualizations of the main effects for Driver Age\footnote{Segment names have been anonymized.} and CRM Coefficient. For each $\mathbf{x}_i \in \mathcal{X}$, the shape function for variable $X_j$ $f_j(\mathbf{x}_i)$ with $j \in \{1, \ldots, p\}$ locally measures its effect for individual $i$. Similarly to the regression coefficients in a GLM model, the values of the average shape functions can be compared to assess the importance of each feature. Note that these shape functions are centered and normalized to unit variance. In addition, the use of the logarithmic link function in the claim models means that the y-axis of shape functions is in logarithmic space.
Each plot is divided into two panels: the upper panel shows the estimated shape function, representing the feature's contribution to the prediction (\textit{Score}), while the lower panel displays a histogram of the feature's distribution (\textit{Exposure}) in the training dataset. Uncertainty in the shape functions is represented by error bars, which account for both data sparsity in different regions of the feature space and variability of the learned model. These uncertainties are estimated using a bagging procedure, where the EBM model is trained on multiple randomly selected subsets of the data. For both Driver Age and CRM Coefficient, we observe a clear and interpretable effect on claim severity. Driver Age generally shows a decreasing effect on predicted severity for younger ages, stabilizing for ages between 40 and 50, as well as between 60 and 70, before increasing for older ages, which is generally consistent with observed claim severity patterns in car insurance. CRM Coefficient exhibits a generally increasing effect on claim severity, indicating that higher CRM values correspond to higher predicted claims.

\begin{figure}[h!]
\centering
\begin{subfigure}{.5\textwidth}
  \centering
  \includegraphics[width=\linewidth]{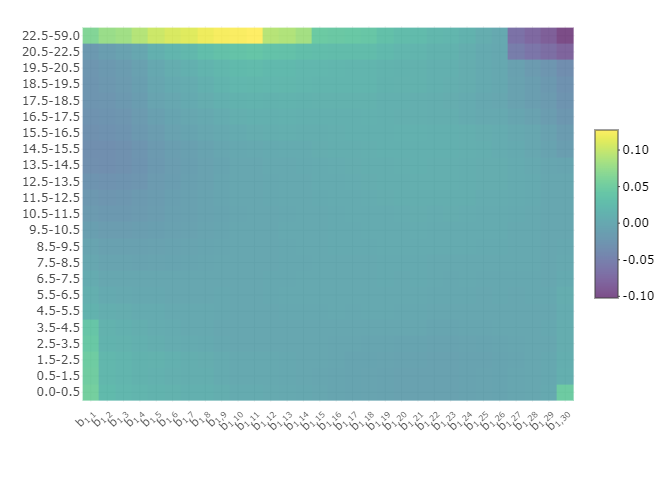}
  \caption{Driver Age $\times$ Vehicle Age}
  \label{fig:heatmap_age_COND_age_VEH_sev}
\end{subfigure}%
\begin{subfigure}{.5\textwidth}
  \centering
  \includegraphics[width=\linewidth]{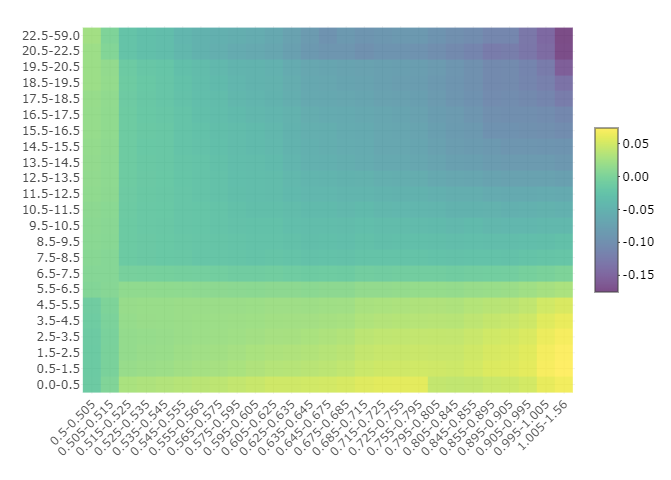}
  \caption{CRM Coefficient $\times$ Vehicle Age}
  \label{fig:heatmap_cof_CRM_age_VEH_sev}
\end{subfigure}%
\caption{Heatmaps of the predicted pairwise interaction terms $f_{(j_1,j_2)}$ in the claim severity EBM$^2$ model. The displayed interactions correspond to (Driver Age, Vehicle Age) in Figure~\ref{fig:heatmap_age_COND_age_VEH_sev} and (CRM Coefficient, Vehicle Age) in Figure~\ref{fig:heatmap_cof_CRM_age_VEH_sev}. Color intensity reflects the magnitude of the interaction effect on predicted claim severity.}
\label{fig:heatmap_sev}
\end{figure}

Heatmaps of the selected pairwise interaction terms identified by the GA$^2$M algorithm are presented in Figure~\ref{fig:heatmap_sev} for the claim severity model. We focus on two illustrative examples in which notable interaction patterns emerge. The interaction between Driver Age and Vehicle Age reveals an increase in severity for very young drivers with very old vehicles, while showing a decrease for older drivers with older vehicles. Next, the interaction between the CRM Coefficient and Vehicle Age indicates that vehicles with a high CRM Coefficient, i.e. those that have experienced a higher claim frequency in the past, and low vehicle age tend to have higher claim severity. However, beyond a certain Vehicle Age, predicted severity appears to decline as the CRM Coefficient increases.

\subsubsection{EBM framework and Shapley values}
\label{subsec:var_ebm_shap}

We conclude this presentation by emphasizing the relationship between EBM predictions (up to the log-link transformation) and SHAP values. Due to its additive structure, each feature in the EBM contributes to the prediction in a modular and separable manner. As discussed in Section~\ref{subsubsec:relation_shapley}, the EBM framework is closely related to the Shapley value and the Shapley interaction index. This structural property ensures that SHAP values and Shapley interaction indices faithfully recover the full set of effects encoded in the model's shape functions.

\begin{figure}[ht!]
\centering
\begin{subfigure}{.5\textwidth}
  \centering
  \includegraphics[width=\linewidth]{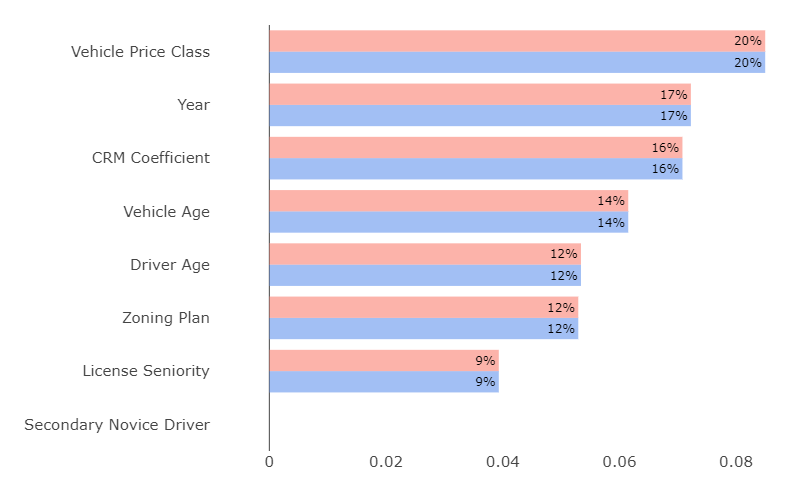}
  \caption{Relative feature importance}
  \label{fig:ebm_shap_imp}
\end{subfigure}%
\begin{subfigure}{.5\textwidth}
  \centering
  \includegraphics[width=\linewidth]{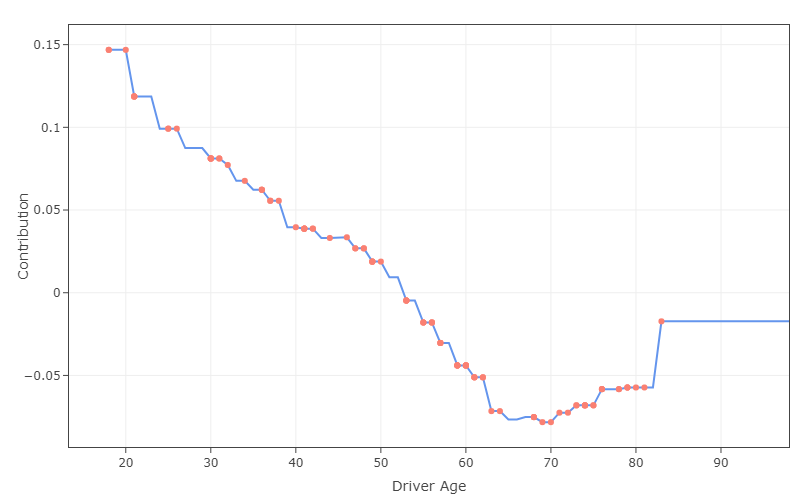}
  \caption{Local contributions for the Driver Age feature}
  \label{fig:driver_age_ebm_shap}
\end{subfigure}
\caption{Comparison of feature contributions from the EBM model (without interactions) and SHAP values on the testing dataset. Relative feature importance and local effects are shown in blue for EBM and in pink for SHAP.}
\label{fig:ebm_shap}
\end{figure}

To illustrate this property, Figure~\ref{fig:ebm_shap} provides an example of recovery for the EBM model without interactions on the testing dataset. We choose the non-interaction model for simplicity, avoiding the computationally expensive calculation of second-order interactions. Figure~\ref{fig:ebm_shap_imp} compares feature contributions computed from the shape functions of this EBM model with the SHAP values estimated using the Kernel SHAP approximation algorithm \citep{lundbergUnifiedApproachInterpreting2017b}. Globally, Figure~\ref{fig:ebm_shap_imp} demonstrates that the relative feature importance, as defined in Eq.~\eqref{eq:feat_importance_ebm}, coincides perfectly with that computed from the SHAP values on the training set. Locally, Figure~\ref{fig:driver_age_ebm_shap} illustrates the same perfect recovery property of the additive EBM model for the feature Driver Age. Analogous results are observed for all other features. It should be noted that minor differences may appear on the test set, due to distributional discrepancies between the background sample used by the Kernel SHAP method and the test set.

\begin{figure}[ht!]
\centering
\begin{subfigure}{.33\textwidth}
  \centering
  \includegraphics[width=\linewidth]{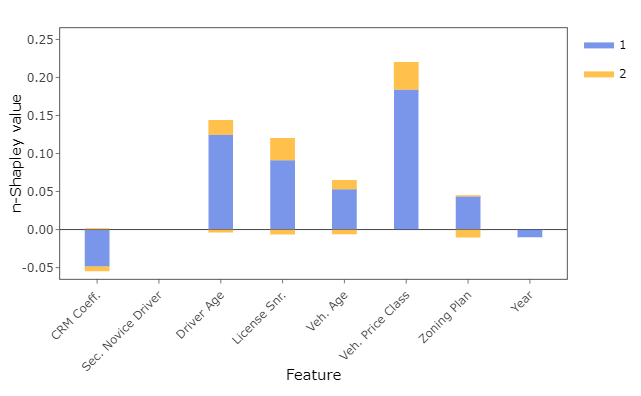}
  \caption{EBM$^2$}
  \label{fig:nshap_EBM}
\end{subfigure}%
\begin{subfigure}{.33\textwidth}
  \centering
  \includegraphics[width=\linewidth]{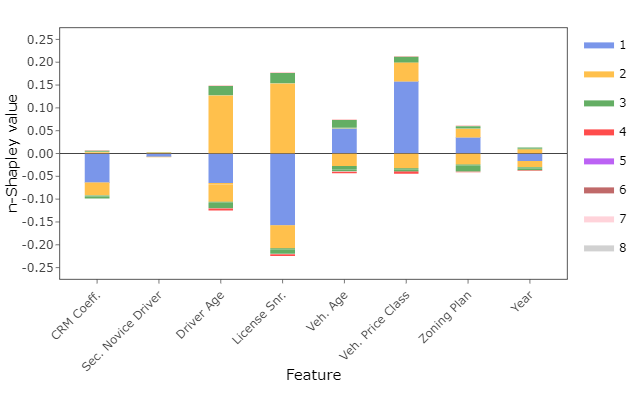}
  \caption{XGB}
  \label{fig:nshap_XGB}
\end{subfigure}
\begin{subfigure}{.33\textwidth}
  \centering
  \includegraphics[width=\linewidth]{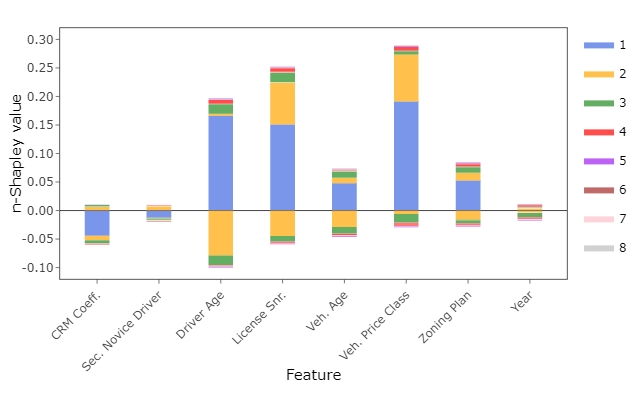}
  \caption{LocalGLMnet}
  \label{fig:nshap_LGLM}
\end{subfigure}
\caption{Comparison of the $n$-Shapley values for the EBM$^2$ (Figure~\ref{fig:nshap_EBM}), XGB (Figure~\ref{fig:nshap_XGB}) and LocalGLMnet (Figure~\ref{fig:nshap_LGLM}) models for a single randomly selected observation.}
\label{fig:nshap}
\end{figure}

Another illustration of this remarkable property of additive models can be obtained by computing the $n$-Shapley values $\Phi_{\mathcal{C}}^n$, as defined in Eq.~\eqref{eq:n_shapley_definition}. In Figure~\ref{fig:nshap}, we compare $\Phi_{\mathcal{C}}^n$ for the EBM$^2$ model (Figure~\ref{fig:nshap_EBM}), the XGB model (Figure~\ref{fig:nshap_XGB}), and the LocalGLMnet (Figure~\ref{fig:nshap_LGLM}). As expected, the contributions in the EBM$^2$ model are restricted to pairwise (order-2) interactions for each feature, perfectly matching the model specification. In contrast, the XGB and LocalGLMnet models allow for higher-order interactions due to the depth of the regression trees and neural networks used during training. Consequently, contributions for some variables can reach order eight, which greatly complicates the interpretability of these models. In practice, such higher-order interactions are rarely evaluated, since practitioners typically focus on standard SHAP values, which capture only main effects (order 1). For the XGB model, considering only main effects can be misleading. Variables such as Driver Age or License Seniority may appear less important than their total contribution, which includes higher-order interactions. Similarly, for the LocalGLMnet model, the ranking of the three most important variables can be misinterpreted if interactions are ignored.

It should be noted however that the fidelity property of additive models does not imply that the recovered effects reflect true data-generating effects or causal relationships. Rather, it guarantees a faithful reconstruction of the effects according to the model specification, which is much harder to achieve with black-box models as illustrated above. Furthermore, the local contributions of EBM$^2$ and other models do not necessarily coincide for each variable, as the computation of these contributions depends on the model used for evaluation.

\section{Conclusion}
\label{sec_ccl}

In this paper, we explore the potential of using the EBM model for analyzing claim frequency and severity in non-life insurance, which are the two key components in determining technical premiums. We develop an analytical framework applied to car insurance data, enabling both a rigorous comparison of predictive performance, reliability, and an assessment of interpretability properties. EBM is a glass-box model based on a GAM structure, where shape functions are independently trained using a cyclic gradient boosting trees procedure. Due to this additive structure, each feature contribution can be directly quantified through its impact on the score function. In addition, the GA$^2$M selection algorithm allows for the identification of the most relevant pairwise interaction terms while preserving interpretability.  

Our empirical results show that EBM achieves competitive out-of-sample predictive performance, particularly when selected pairwise interactions are incorporated. Across multiple evaluation metrics, including deviance-based comparisons and dominance tests, EBM and EBM$^2$ remain competitive with state-of-the-art machine learning models used in actuarial literature. At the same time, the additive structure of EBM ensures a faithful and transparent decomposition of predictions into main and interaction effects, both at the global and local levels. In contrast to black-box models, where higher-order interactions may remain hidden and difficult to assess, EBM guarantees consistency between model specification and interpretability outputs. From a practical standpoint, this framework is especially valuable in insurance pricing. It enables transparent validation of feature effects, facilitates interaction screening, and provides a principled basis for variable selection. Moreover, EBM can serve as a powerful intermediary tool for constructing high-performing and fully interpretable pricing models, such as enhanced GLMs, by guiding feature engineering and the integration of selected bivariate effects. Unlike black-box approaches such as GBM or DL models, EBM reduces the need for complex surrogate modeling strategies to recover interpretability ex post.

Several avenues for future research emerge from this work. First, EBM training currently relies on deviance as the optimization criterion. Exploring alternative loss functions specifically tailored to insurance pricing objectives may further improve its performance. Second, dimension-wise early stopping strategies, as proposed by \citet{zakrisson_tree-based_2025}, could be adapted to enhance model training. Finally, it would be interesting to investigate the relevance of this model for more complex datasets, such as those with spatial interactions or temporal components, as well as other risks, such as home insurance or health insurance.

\paragraph{Acknowledgments}

The authors are very grateful to the Editor-in-Chief and the two anonymous referees for their constructive suggestions, which led to significant improvements in this article. The authors would like to thank Paul Koch (Microsoft) for his valuable advice on the use of the EBM algorithm. Quentin Guibert gratefully acknowledges funding from the Institut Europlace de Finance (IEF) as part of the research project "Machine Learning and Explainability". This version of the article has been accepted for publication, after peer review but is
not the Version of Record and does not reflect post-acceptance improvements, or any corrections. The Version of Record is
available online at: \url{http://dx.doi.org/10.1007/s13385-026-00461-y}.

\paragraph{Supplementary material}
The results in this paper are obtained using $\mathtt{R}$ and $\mathtt{Python}$. Supplementary material related to this paper can be found at \url{https://github.com/MarketaKrupova/EBM-Car-Insurance}.

\paragraph{Declarations}
The authors declare no potential conflict of interests.

\paragraph{Open Access}
Open Access This article is licensed under a Creative Commons Attribution 4.0 International License, which permits use, sharing, adaptation, distribution and reproduction in any medium or format, as long as you give appropriate credit to the original author(s) and the source, provide a link to the Creative Commons license, and indicate if changes were made. The images or other third-party material in this article are included in the article’s Creative Commons license, unless indicated otherwise in a credit line to the material. If material is not included in the article’s Creative Commons license and your intended use is not permitted by statutory regulation or exceeds the permitted use, you will need to obtain permission directly from the copyright holder. To view a copy of this license, visit http://creativecommons.org/ licenses/by/4.0/.


\addcontentsline{toc}{section}{References}

\printbibliography


\begin{appendices}

\input{appendix_v2.0}

\end{appendices}

\end{document}

%% file: appendix_v2.0.tex
\section{Data description}
\label{sec:app_data}

Table~\ref{tab:features_modeling} describes all the variables of this dataset, as well as their use for claim frequency and severity applications.  The selected features can be schematically divided into four categories:
\begin{enumerate}
    \item Characteristics of the policyholder (Driver Age, License Seniority, Secondary Novice Driver, CRM Coefficient\footnote{French system that rewards or penalizes drivers based on their driving history. The initial coefficient is 1 and for each year without accident, the coefficient is reduced by 5\%. After 13 years without accident, the coefficient reaches 0.5 and reduces the initial premium by 50\%. Conversely, in the event of an accident, the coefficient is increased depending on the severity of the accident and the driver's responsibility and penalizes the initial premium.}),
    \item Characteristics of the vehicle (Vehicle Price Class, Vehicle Age, Vehicle Use),
    \item Geographical zone (Zoning Plan),
    \item Temporality (Year).  
\end{enumerate}

\begin{table}[!ht]
    \caption{Features selected for the claim frequency and severity modeling of the full accidental damage cover. \label{tab:features_modeling}}
    \fontsize{8}{10}\selectfont
    \centering
    \renewcommand{\arraystretch}{1.3} 
    \setlength{\tabcolsep}{6pt}
    \begin{tabular} {cccc}
    \toprule
        \textbf{Feature} & \textbf{Description} & \textbf{Frequency} & \textbf{Severity} \\ \midrule
        Driver Age & Age of the main driver in years (from ages 17 to 100) & \checkmark & \checkmark \\
        License Seniority & License seniority of the main driver in years (from 0 to 82 years) & \checkmark & \checkmark \\
        Secondary Novice Driver & Presence of a secondary novice driver & \checkmark & \checkmark \\
        CRM Coefficient & Reduction-increase (bonus-malus) coefficient & \checkmark & \checkmark \\
        Vehicle Price Class & Cover-specific vehicle price class & \checkmark & \checkmark \\
        Vehicle Age & Age of the vehicle in years & $\times$ & \checkmark \\
        Vehicle Use & Main use of the vehicle & \checkmark & $\times$ \\
        Zoning Plan & Cover-specific zoning plan & \checkmark & \checkmark \\
        Year & Calendar year & \checkmark & \checkmark \\
    \bottomrule 
    \end{tabular}    
\end{table}

\begin{figure}[!ht]
\centering
\begin{subfigure}{.24\textwidth}
  \centering
  \includegraphics[width=\linewidth]{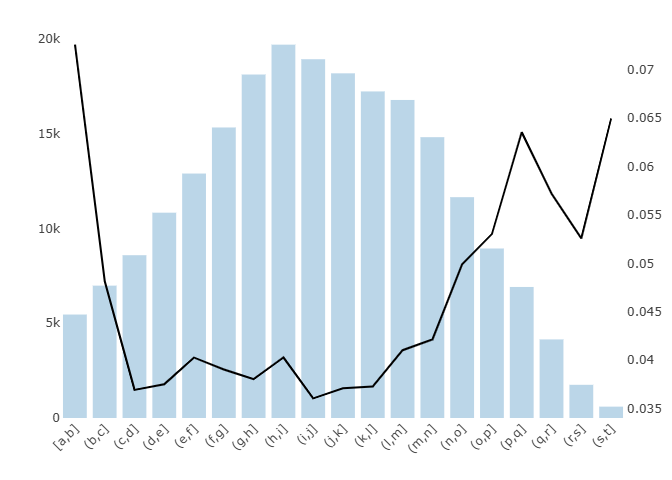}
  \caption{Driver Age}
  \label{fig:age_COND_freq}
\end{subfigure}%
\begin{subfigure}{.24\textwidth}
  \centering
  \includegraphics[width=\linewidth]{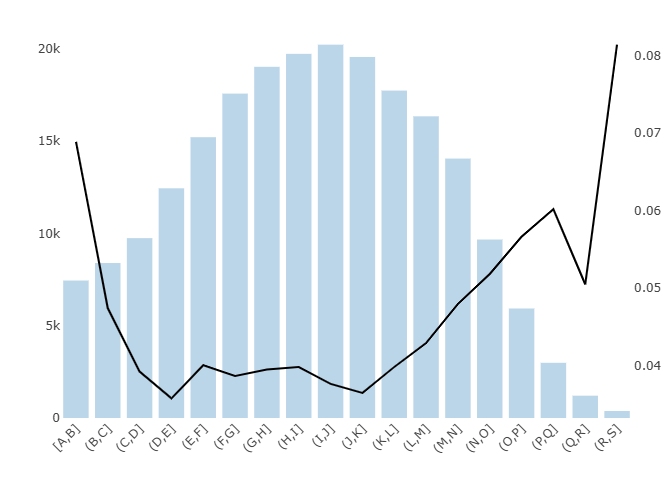}
  \caption{License Seniority}
  \label{fig:anc_COND_freq}
\end{subfigure}
\begin{subfigure}{.24\textwidth}
  \centering
  \includegraphics[width=\linewidth]{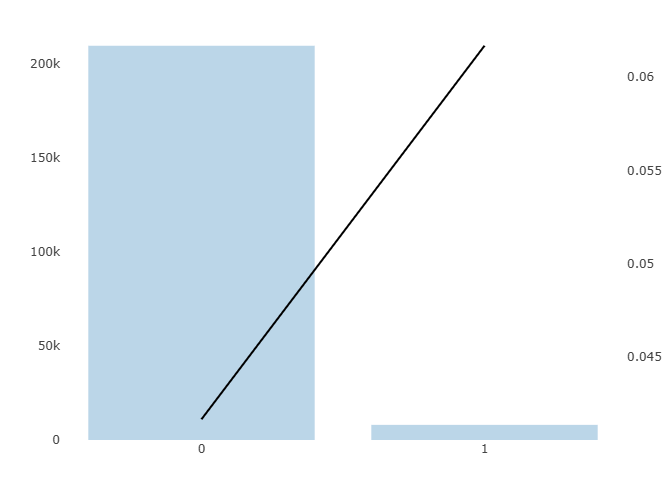}
  \caption{Sec. Novice Driver}
  \label{fig:novice_COND2_freq}
\end{subfigure}%
\begin{subfigure}{.24\textwidth}
  \centering
  \includegraphics[width=\linewidth]{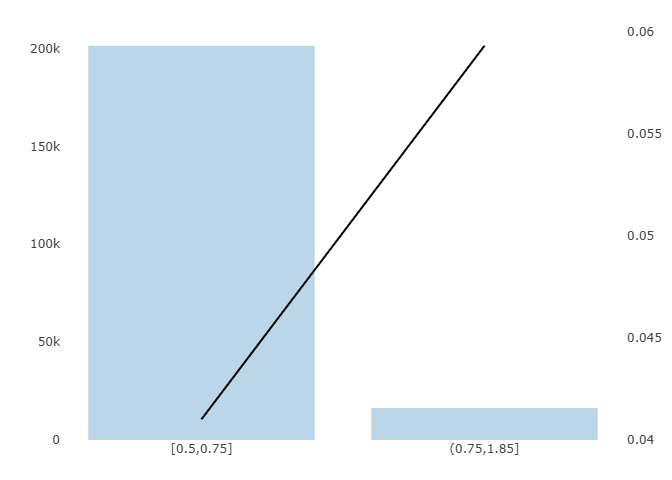}
  \caption{CRM Coefficient}
  \label{fig:cof_CRM_freq}
\end{subfigure}
\begin{subfigure}{.24\textwidth}
  \centering
  \includegraphics[width=\linewidth]{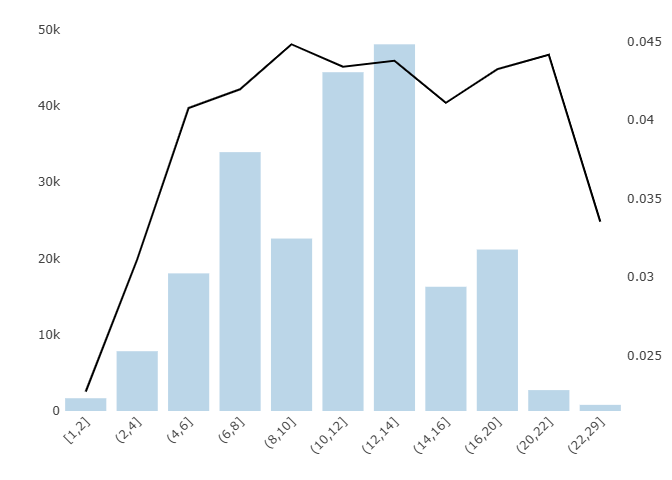}
  \caption{Vehicle Price Class}
  \label{fig:prix_SRA_freq}
\end{subfigure}%
\begin{subfigure}{.24\textwidth}
  \centering
  \includegraphics[width=\linewidth]{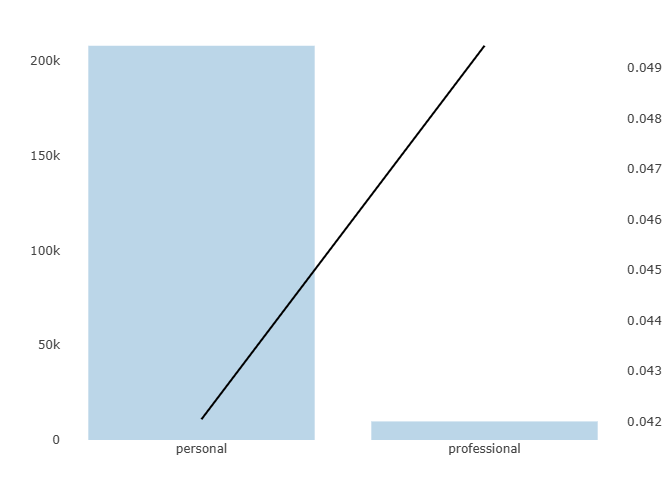}
  \caption{Vehicle Use}
  \label{fig:usage_freq}
\end{subfigure}
\begin{subfigure}{.24\textwidth}
  \centering
  \includegraphics[width=\linewidth]{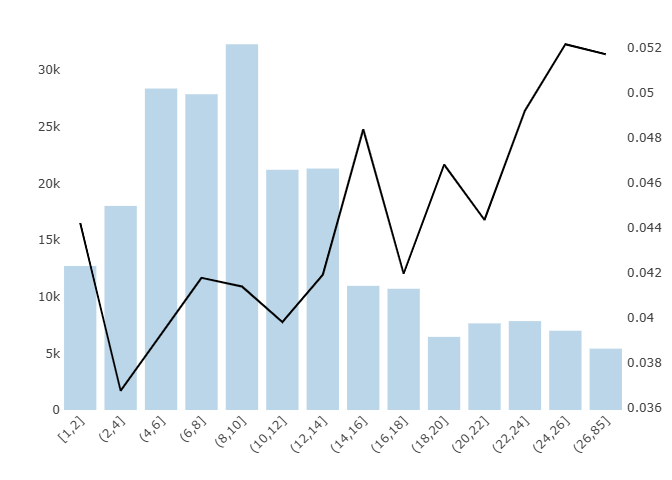}
  \caption{Zoning Plan}
  \label{fig:zonier_DTA_freq}
\end{subfigure}%
\begin{subfigure}{.24\textwidth}
  \centering
  \includegraphics[width=\linewidth]{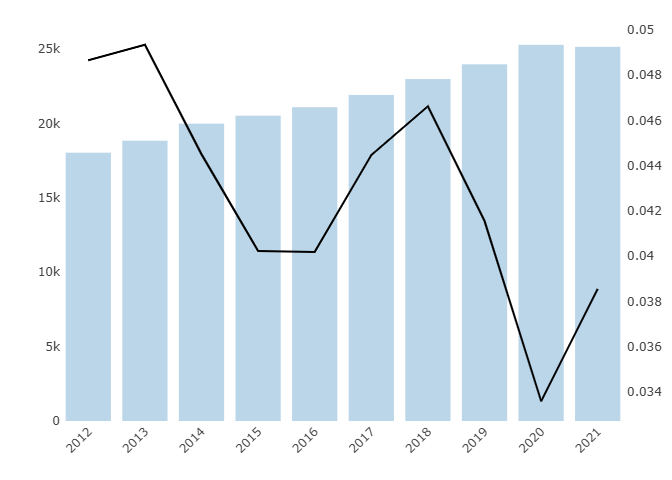}
  \caption{Year}
  \label{fig:annee_freq}
\end{subfigure}
\caption{Univariate analysis of features for claim frequency. For each feature, the bar chart shows the exposure-to-risk in thousands (left y-axis) and the line chart displays the empirical claim frequency (right y-axis).}
\label{fig:univariate_freq}
\end{figure}

\begin{figure}[!ht]
\centering
\begin{subfigure}{.24\textwidth}
  \centering
  \includegraphics[width=\linewidth]{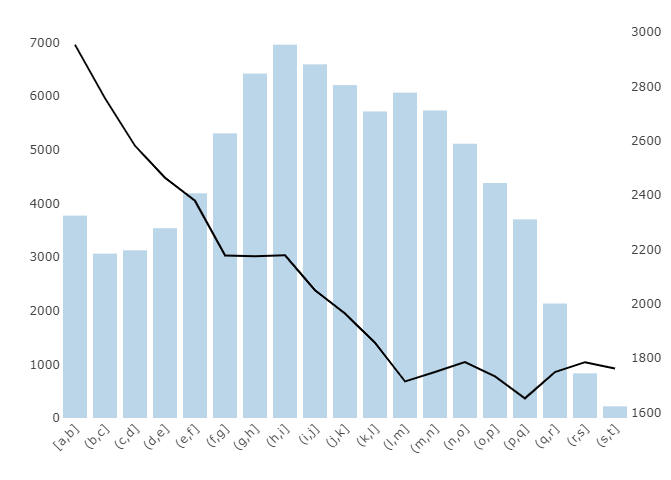}
  \caption{Driver Age}
  \label{fig:age_COND_sev}
\end{subfigure}%
\begin{subfigure}{.24\textwidth}
  \centering
  \includegraphics[width=\linewidth]{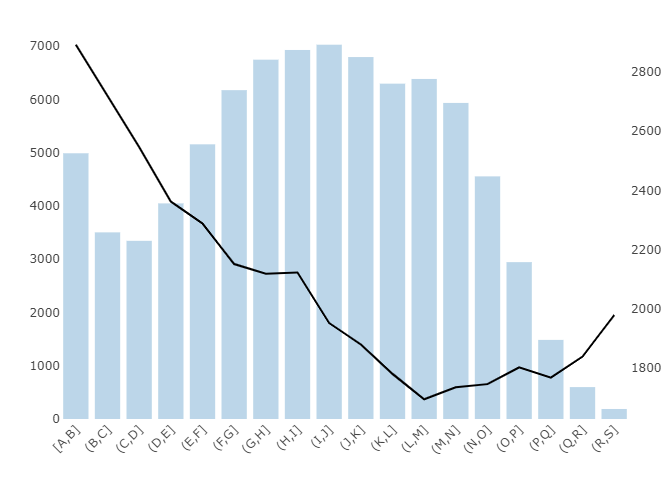}
  \caption{License Seniority}
  \label{fig:anc_COND_sev}
\end{subfigure}
\begin{subfigure}{.24\textwidth}
  \centering
  \includegraphics[width=\linewidth]{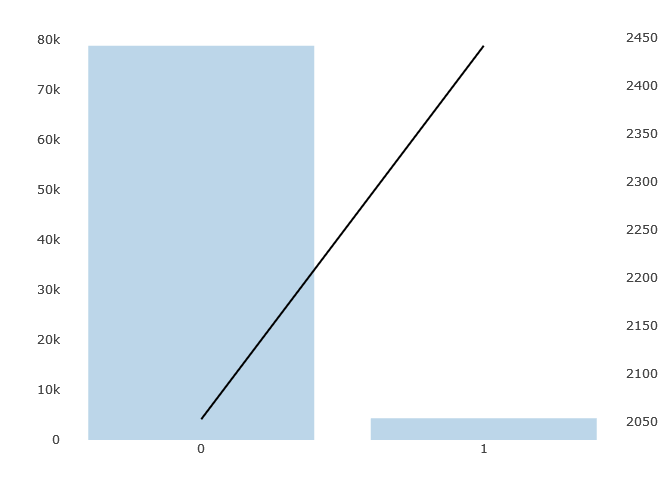}
  \caption{Sec. Novice Driver}
  \label{fig:novice_COND2_sev}
\end{subfigure}%
\begin{subfigure}{.24\textwidth}
  \centering
  \includegraphics[width=\linewidth]{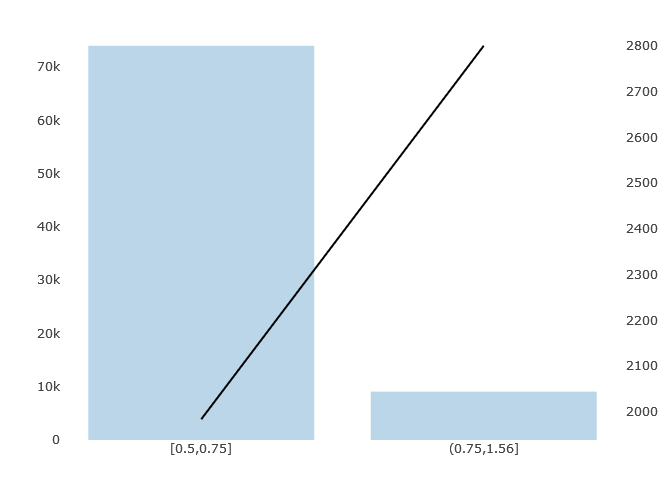}
  \caption{CRM Coefficient}
  \label{fig:cof_CRM_sev}
\end{subfigure}
\begin{subfigure}{.24\textwidth}
  \centering
  \includegraphics[width=\linewidth]{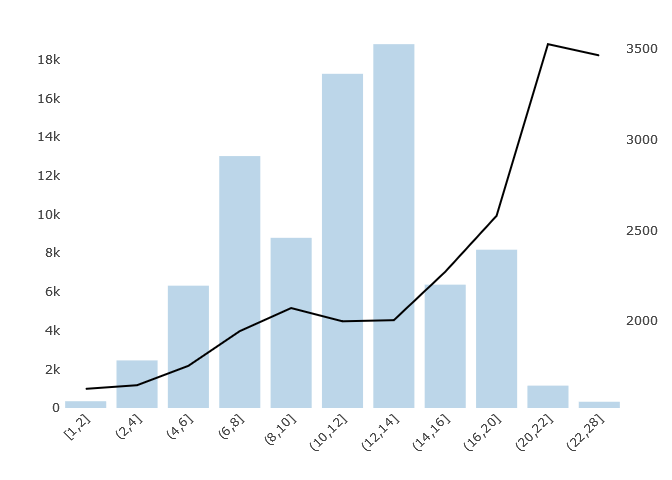}
  \caption{Vehicle Price Class}
  \label{fig:prix_SRA_sev}
\end{subfigure}%
\begin{subfigure}{.24\textwidth}
  \centering
  \includegraphics[width=\linewidth]{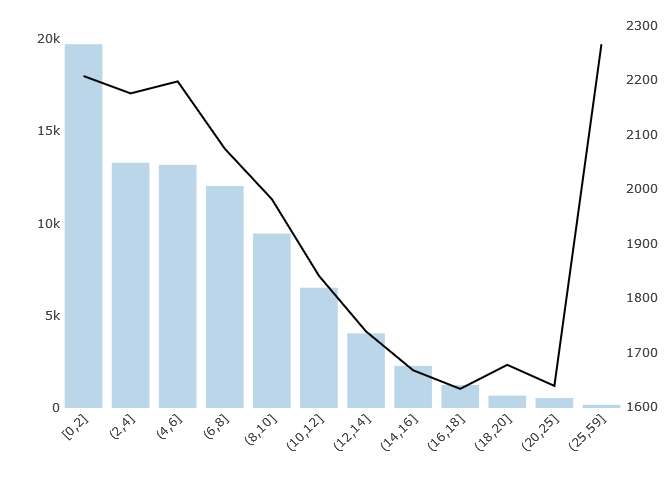}
  \caption{Vehicle Age}
  \label{fig:age_VEH_sev}
\end{subfigure}
\begin{subfigure}{.24\textwidth}
  \centering
  \includegraphics[width=\linewidth]{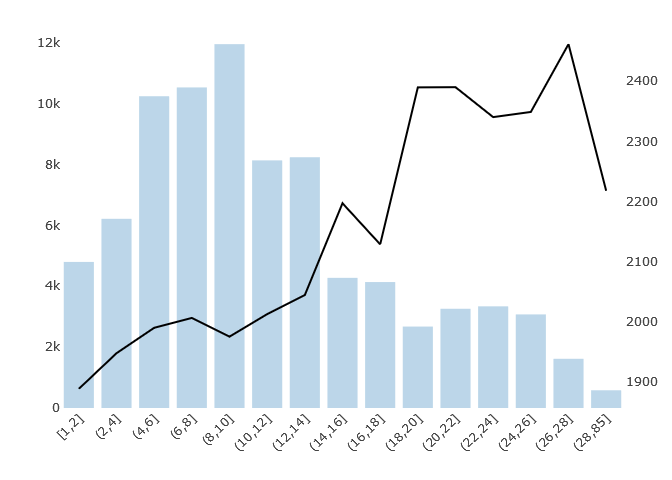}
  \caption{Zoning Plan}
  \label{fig:zonier_DTA_sev}
\end{subfigure}%
\begin{subfigure}{.24\textwidth}
  \centering
  \includegraphics[width=\linewidth]{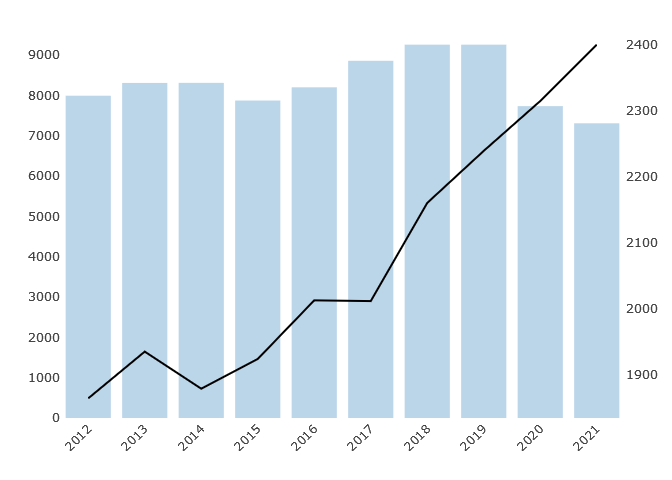}
  \caption{Year}
  \label{fig:annee_sev}
\end{subfigure}
\caption{Univariate analysis of features for claim severity. For each feature, the bar chart shows the number of claims in thousands (left y-axis) and the line chart displays the empirical average claim cost (right y-axis).}
\label{fig:univariate_sev}
\end{figure}

Figure~\ref{fig:univariate_freq} presents a distribution histogram of each feature included in the claim frequency dataset, along with the empirical claim frequency. Figure~\ref{fig:univariate_sev} displays a distribution histogram of each feature included in the claim severity dataset, along with the evolution of the empirical average claim cost.

\section{Implementation EBM with \textbf{InterpretML}}
\label{sec:app_ebm_method}

\textbf{InterpretML} is the first open-source \texttt{Python} package to include an implementation of EBM \citep{nori_interpretml_2019}. In this appendix, we specify the hyperparameters of this algorithm and explain how to account for interactions between features. The user can control the learning procedure of the algorithm by placing the cursor on the following elements of Table~\ref{tab:EBM_parameters}:

\begin{itemize}
    \item GAM structure by selecting the main effect features and interaction features as well as the loss function to optimize (in yellow).
    \item Tree structure by controlling the complexity of the decision trees used as building blocks (in green).
    \item Binning procedure by choosing the number of bins (in red).
    \item Cyclic boosting procedure (in blue).
\end{itemize}

\begin{table}[ht]
	\caption{List of hyperparameters for the \textit{ExplainableBoostingRegressor} algorithm in \textbf{InterpretML}, see details in the official documentation \url{https://interpret.ml/docs/hyperparameters.html} \label{tab:EBM_parameters}.}
    \fontsize{8}{10}\selectfont
    \centering
    \begin{tabular}{p{4cm}p{1.5cm}p{8cm}} \toprule
        \textbf{Parameter} & \textbf{Default} & \textbf{Explanation} \\ \midrule
        \rowcolor{LightYellow}
        \texttt{feature\_names} & None & List of feature names \\ 
        \rowcolor{LightYellow}
        \texttt{feature\_types} & None & List of feature types \\ 
        \rowcolor{LightYellow}
        \texttt{exclude} & None & Features to be excluded \\ 
        \rowcolor{LightYellow}
        \texttt{interactions} & 0.9 & Interaction terms to be included \\ 
        \rowcolor{LightYellow}
        \texttt{objective} & RMSE & Objective to optimize \\ 
        \rowcolor{LightGreen}
        \texttt{min\_samples\_leaf} & 2 & Minimum number of samples in the leaves \\ 
        \rowcolor{LightGreen}
        \texttt{min\_hessian} & 0.0001 & Minimum hessian required to consider a potential split valid \\ 
        \rowcolor{LightGreen}
        \texttt{max\_leaves} & 3 & Maximum number of leaves in each tree \\ 
        \rowcolor{LightRed}
        \texttt{max\_bins} & 1024 & Maximum number of bins per feature \\ 
        \rowcolor{LightRed}
        \texttt{max\_interaction\_bins} & 32 & Maximum number of bins per feature for interaction terms \\ 
        \rowcolor{LightCyan}
        \texttt{max\_rounds} & 25000 & Number of iterations \\ 
        \rowcolor{LightCyan}
        \texttt{smoothing\_rounds} & 200 & Number of initial highly regularized iterations for main effect feature graphs \\ 
        \rowcolor{LightCyan}
        \texttt{interaction\_smoothing\_rounds} & 50 & Number of initial highly regularized iterations for interaction feature graphs \\ 
        \rowcolor{LightCyan}
        \texttt{early\_stopping\_rounds} & 50 & Number of iterations of no improvement to turn on early stopping \\ 
        \rowcolor{LightCyan}
        \texttt{early\_stopping\_tolerance} & 0.00001 & Tolerance to prevent early stopping \\ 
        \rowcolor{LightCyan}
        \texttt{validation\_size} & 0.15 & Size of the validation set for early stopping and creation of outer bags \\ 
        \rowcolor{LightCyan}
        \texttt{outer\_bags} & 14 & Number of outer bags for generating error bounds and adding smoothness to graphs \\ 
        \rowcolor{LightCyan}
        \texttt{inner\_bags} & 0 & Subsamples drawn with replacement at the time of growing individual trees during boosting \\ 
        \rowcolor{LightCyan}
        \texttt{learning\_rate} & 0.01 & Learning rate for boosting \\ 
        \rowcolor{LightCyan}
        \texttt{greedy\_ratio} & 1.5 & Proportion of greedy boosting steps relative to cyclic boosting steps \\ 
        \rowcolor{LightCyan}
        \texttt{cyclic\_progress} & 1 & Proportion of boosting cycles actively improving model's performance \\ \bottomrule
    \end{tabular}%
\end{table}

In our applications, the Poisson deviance (resp. the gamma deviance) can be implemented with EBM by assigning $\texttt{objective} = \texttt{poisson\_deviance}$ (resp. $\texttt{objective} = \texttt{gamma\_deviance}$). Offsetting exposure-to-risk can be achieved during the fitting procedure with the \texttt{init\_score} parameter, a per-sample initialization score. Exposure-to-risk is then added to the final score at prediction time. For both EBM gamma and Poisson, the automatically assigned link function $g$ is the logarithmic link function, which is often used in insurance pricing applications.

Interaction terms to be included in the model can be passed on to the \texttt{interactions} parameter in three different ways. An integer specifies the count of interactions to be automatically selected, a percentage determines the integer count of interactions by multiplying the number of features by this percentage, or a list of tuples explicitly gives the features to be considered as interaction terms. Monotonic constraints for each feature's relationship with the target may be specified during the fitting process using \texttt{monotone\_constraints} parameter or ex post using \texttt{monotonize} function based on isotonic regression. The \textbf{InterpretML} package also offers a visualization platform for global and local explanations obtained with EBM.

\section{Hyperparameters tuning and model specifications}
\label{sec:app_model_spec}

This section complements Section~\ref{subsec:appli_benchmark_models} by providing additional methodological details regarding the training and fine-tuning strategy (Appendix~\ref{subsec:appli_train_tuning}), as well as the implementation of the different models considered in the study (Appendix~\ref{subsec:app_model_spec}). 

\subsection{Training and fine tuning procedures}
\label{subsec:appli_train_tuning}

The training strategy adopted for all models described in Section~\ref{subsec:appli_benchmark_models} is based on a standard hold-out scheme, with a split into training and testing datasets, denoted $\mathcal{D}^{\text{train}}_{g}$ and $\mathcal{D}^{\text{test}}_{g}$ for $g \in \{\text{freq}, \text{sev}\}$, corresponding to the frequency and severity datasets, respectively. More precisely, 80\% of the observations are used for model estimation, while the remaining 20\% are reserved for the assessment of out-of-sample predictive performance. To ensure comparability and mitigate potential distributional shifts, the partition is constructed so that both subsets display similar empirical distributions across all explanatory variables. The deviance, defined in Equation~\eqref{eq:deviance}, is used as the loss function. More precisely, following \citet{henckaerts_boosting_2021}, we adopt the Poisson deviance associated with a GLM framework as the loss function for modeling claim frequency
\begin{equation} \label{eq:deviance_pois}
L_{\text{freq}}(\mathbf{y}, F(\mathbf{x})) = \frac{2}{n} \sum_{i=1}^{n}{y_i \ln{\left( \frac{y_i}{e_i F(\mathbf{x}_i)}\right)} -  (y_i - e_i F(\mathbf{x}_i))},
\quad y_i \in \mathcal{Y},
\quad \mathbf{x}_i \in \mathcal{X}.
\end{equation}
where $e_i$ the exposure-to-risk for policyholder $i$.
Similarly for modeling claim severity, we consider the deviance of a gamma GLM as the loss function  
\begin{equation} \label{eq:deviance_gamma}
L_{\text{sev}}(\mathbf{y}, F(\mathbf{x})) = \frac{2}{n} \sum_{i=1}^{n}{\alpha_i \left(\frac{y_i -  F(\mathbf{x}_i)}{F(\mathbf{x}_i)} - \ln{\left( \frac{y_i}{ F(\mathbf{x}_i)}\right)}\right)},
\quad y_i \in \mathcal{Y},
\quad \mathbf{x}_i \in \mathcal{X}.
\end{equation}
where $\alpha_i$ corresponds to a shape parameter that we ignore in practice for a point prediction model.

Regarding the distributional models GAMBoostLSS and cyc-GBM, we model the scale and shape parameters using the gamma deviance loss function for claim severity. For claim frequency, we assume a Negative Binomial type 2 (NB2) distribution and regress both the location and dispersion parameters by minimizing the corresponding NB2 deviance loss.

Hyperparameter tuning is performed exclusively on the training set using a three-fold cross-validation procedure. Specifically, $\mathcal{D}^{\text{train}}_{g}$ is divided into three mutually exclusive folds of approximately equal size. At each iteration, the model is trained on two folds and validated on the remaining one. Once the optimal hyperparameters have been selected, the model is finally re-estimated on the entire training set $\mathcal{D}^{\text{train}}_{g}$ and subsequently evaluated on the test set $\mathcal{D}^{\text{test}}_{g}$. The ranges and candidate values of the hyperparameters considered for each model are reported in Appendix~\ref{subsec:app_model_spec}.

The search for optimal hyperparameters is carried out using two optimization strategies. By default, we implement a grid search procedure over a predefined hyperparameter space, exhaustively evaluating all candidate configurations and selecting the model that minimizes the loss function. However, this approach becomes computationally prohibitive as the dimension of the hyperparameter space increases. For models relying on neural network architectures, such as NAM and LocalGLMNet described in Appendix~\ref{subsec:app_model_spec}, we instead employ Bayesian optimization \citep{nogueira2014}, which provides a more computationally efficient strategy for hyperparameter tuning.

\subsection{Model specifications}
\label{subsec:app_model_spec}

This appendix provides the detailed specifications of the competing models to ensure reproducibility of the empirical evaluation. The models are considered for both claim severity and claim frequency.

\subsubsection{GLM bins}
\label{subsubsec:app_glm}

To construct a GLM with predictive performance comparable to the other models considered, we follow an approach similar to the surrogate GLM framework proposed by \citet{henckaerts_when_2022}. The idea consists in extracting structural information from a predictive model through its partial dependence (PD) functions. To this end, we rely on the EBM model without interactions, for which the partial dependence functions \citep{friedman_greedy_2001} can be computed directly due to its additive structure. 

For a given predictive model $F$, the univariate partial dependence (PD) function associated with variable $X_j$ is defined as
\begin{equation}\label{eq:pdp}
PD_j(x_j^\star) = \esp{F(x_j^\star,  \mathbf{\Xvec}_{-j})},
\end{equation}
where the expectation is taken with respect to the marginal distribution of $\mathbf{X}_{-j}$, i.e. the vector of covariates excluding $X_j$. In this formula, the value of $X_j$ in $\Xvec$ is replaced by $x_j^\star \in \mathcal{X}_j$, where $\mathcal{X}_j$ denotes the support of variable $X_j$.
For an EBM model without interaction terms, we can apply the link function $g$ to Eq.~\eqref{eq:pdp}. Using Eq.~\eqref{eq:predict_gam}, this yields
\begin{equation}\label{eq:pdp_ebm}
g\left(PD_j(x_j^\star)\right) = f_j(x_j^\star) + C,
\end{equation}
where $C$ is a constant that does not depend on $x_j^\star$. This observation allows us to apply the \texttt{maidrr} algorithm of \citet{henckaerts_when_2022}[Algorithm 1] without considering the interaction selection step for simplicity. The algorithm is based on a clustering procedure applied to values where the PD values are similar. In the case of an EBM without interactions, this explicitly corresponds to the values where the term $f_j(x_j^\star)$ in Eq.~\eqref{eq:pdp_ebm} is constant, i.e. within the bins defined prior to applying the cyclic boosting procedure described in Section~\ref{subsubsec:fitting_ebm}. This procedure enables feature engineering by transforming the original variables into categorical variables learned from the EBM model. A surrogate GLM is then trained using these newly derived categorical features.

\subsubsection{GAM}
\label{subsubsec:app_gam}

After variable selection, the GAM model is fitted with smoothing parameters estimated using the Generalized Cross Validation (GCV) criterion, as implemented in the \textbf{mgcv} \texttt{R} package \citep{wood_mgcv_2023}.

\subsubsection{GAMboost}
\label{subsubsec:app_gamboost}

GAMboost models are implemented using the \textbf{mboost} \texttt{R} package \citep{hofnerModelbasedBoostingHandson2014}. The Poisson deviance Eq.~\eqref{eq:deviance_pois} is used as the loss function for the claim frequency model, while the gamma deviance Eq.~\eqref{eq:deviance_gamma} is used for the claim severity model. Boosting hyperparameters, including the number of boosting iterations (\texttt{mstop}), the learning rate (\texttt{nu}), and the parameters controlling the tree structure (\texttt{maxdepth}, \texttt{minbucket}, \texttt{minsplit}), are optimized via the grid search described in Table~\ref{tab:grid_gamboost}.
Following the
recommendation in the documentation of the \textbf{mboost} package, the
optimal stopping number of boosting iterations is selected using the \texttt{cvrisk()} function by 3-fold cross-validation.

For the claim frequency model, the search for appropriate hyperparameter values is challenging. In particular, the initial grid considered for \texttt{mstop}, namely $[200, 500, 1000]$, yielded suboptimal results.
Increasing the number of boosting iterations improves model performance but substantially increases computational time. Consequently, the hyperparameter optimization is conducted in two stages. First, the learning rate \texttt{nu} is optimized via grid search while fixing $\texttt{mstop} = 5000$ and keeping all the other parameters at their default values. Second, using the optimal value of \texttt{nu}, the \texttt{maxdepth} parameter is selected, again with $\texttt{mstop} = 5000$ and the remaining parameters set to their default values.
Once these hyperparameters have been calibrated, they are held
fixed and used to train the final model on the training sample, which is
then evaluated on the test set.

\begin{table}[ht]
    \caption{Grid search and optimal hyperparameters for the claim severity and frequency GAMboost models.}
    \fontsize{8}{10}\selectfont
    \centering
    \renewcommand{\arraystretch}{1.3} 
    \setlength{\tabcolsep}{8pt}
    \begin{tabular} {lccccc}
    \toprule
        \textbf{Hyperparameter} & \textbf{Grid Severity} & \textbf{Optimal Severity} & \textbf{Grid Frequency} & \textbf{Optimal Frequency} \\ \midrule
        \texttt{mstop} & $[200, 500, 1000]$ & 200 & 5000 & 5000 \\ 
        \texttt{nu} & $[0.001, 0.01, 0.05, 0.1]$ & 0.1 & $[0.001, 0.01, 0.05, 0.1]$ & 0.1 \\ 
        \texttt{maxdepth} & $[1, 2, 3, 4, 5]$ & 5 & $[1, 3, 5]$ & 3 \\ 
        \texttt{minbucket} & $[5, 10, 20]$ & 5 & 7 & 7 \\ 
        \texttt{minsplit} & $[10, 20]$ & 10 & 20 & 20 \\ \bottomrule
    \end{tabular}
    \label{tab:grid_gamboost}
\end{table}

\subsubsection{GAMboostLSS}
\label{subsubsec:app_gamboostlss}

The GAMboostLSS models are implemented using the \textbf{gamboostLSS} \texttt{R} package \citep{hofner_gamboostlss_2016} and are closely related to the GAMboost implementation. Due to the distributional nature of GAMboostLSS, the NB2 deviance is used as the loss function for the claim frequency model, while the gamma deviance is employed for the claim severity model. Hyperparameter optimization for the claim frequency model follows the same procedure described for GAMboost in Section \ref{subsubsec:app_gamboost}.

\begin{table}[ht]
    \caption{Grid search and optimal hyperparameters for the claim severity and frequency GAMboostLSS models.}
    \fontsize{8}{10}\selectfont
    \centering
    \renewcommand{\arraystretch}{1.3} 
    \setlength{\tabcolsep}{8pt}
    \begin{tabular} {lccccc}
    \toprule
        \textbf{Hyperparameter} & \textbf{Grid Severity} & \textbf{Optimal Severity} & \textbf{Grid Frequency} & \textbf{Optimal Frequency} \\ \midrule
        \texttt{mstop} for \texttt{mu} & $[200, 500, 1000]$ & 200 & 5000 & 5000 \\ 
        \texttt{mstop} for \texttt{sigma} & 200 & 200 & 1000 & 1000 \\ 
        \texttt{nu} & $[0.001, 0.01, 0.05, 0.1]$ & 0.1 & $[0.001, 0.01, 0.05, 0.1]$ & 0.1 \\ 
        \texttt{maxdepth} & $[1, 2, 3, 4, 5]$ & 2 & $[1, 3, 5]$ & 5 \\ 
        \texttt{minbucket} & $[5, 10, 20]$ & 5 & 7 & 7 \\ 
        \texttt{minsplit} & $[10, 20]$ & 10 & 20 & 20 \\ \bottomrule
    \end{tabular}
    \label{tab:grid_gamboostlss}
\end{table}

\subsubsection{cyc-GBM}
\label{subsubsec:appgbm}

For the implementation of the cyclic multi-parameter GBM, we use the \texttt{Python} package \textbf{cyc-gbm} \citep{delong_cyclic_2023}, employing the NB2 deviance as the loss function for the claim frequency model and the gamma deviance for the claim severity model. The boosting parameters (\texttt{n\_estimators} and \texttt{learning\_rate}) and tree-structure parameters (\texttt{max\_depth}, \texttt{min\_samples\_leaf}, and \texttt{min\_samples\_split}) are optimized via the grid search outlined in Table~\ref{tab:grid_gbm}. Due to the high computational cost, the hyperparameter optimization is conducted in two stages. First, the learning rate and the number of estimators are optimized via grid search while keeping all the other parameters at their default values. Second, conditional on the optimal boosting parameters, the tree structure is determined by tuning \texttt{max\_depth}, \texttt{min\_samples\_leaf}, and \texttt{min\_samples\_split}.

Note that an early stopping procedure is proposed by
\citet[Algorithm~3]{delong_cyclic_2023}. However, this
procedure is extremely expensive computationally on our datasets.
Following \citet{chevalierPointProbabilisticGradient2025a}, we therefore
simply determined the optimal number of boosting iterations by 3-fold
cross-validation through a grid search.

\begin{table}[ht]
    \caption{Grid search and optimal hyperparameters for the claim severity and frequency cyc-GBM models.}
    \fontsize{8}{10}\selectfont
    \centering
    \renewcommand{\arraystretch}{1.3} 
    \setlength{\tabcolsep}{8pt}
    \begin{tabular} {lccccc}
    \toprule
        \textbf{Hyperparameter} & \textbf{Grid} & \textbf{Optimal Severity} & \textbf{Optimal Frequency} \\ \midrule
        \texttt{n\_estimators} & $[100, 200, 500, 1000]$ & 100 & 100 \\ 
        \texttt{learning\_rate} & $[0.001, 0.01, 0.05, 0.1]$ & 0.1 & 0.1 \\ 
        \texttt{max\_depth} & $[1, 2, 3]$ & 3 & 3 \\ 
        \texttt{min\_samples\_leaf} & $[1, 5, 10]$ & 5 & 5 \\ 
        \texttt{min\_samples\_split} & $[2, 10, 20]$ & 10 & 2 \\ \bottomrule
    \end{tabular}
    \label{tab:grid_gbm}
\end{table}

\subsubsection{EBM and EBM$^2$}
\label{subsubsec:app_ebm}

The implementation of the EBM model, using the \textbf{InterpretML} \texttt{Python} package, is described in Section~\ref{sec:app_ebm_method}. The hyperparameters optimized via grid search for the claim frequency and severity models are reported in Table~\ref{tab:grid_ebm}.

We remark that the number of boosting rounds is not included in
the cross-validation grid. As reported in Table~\ref{tab:EBM_parameters} and discussed in Remark~\ref{rk:early_stopping_criterion}, EBM sets a maximum number
of boosting iterations (\texttt{max\_rounds}) and relies on an internal
algorithm that determines a global early-stopping criterion. Therefore, early
stopping is not included in our cross-validation grid. At each fit, the
estimation algorithm holds out an internal validation partition from the data it
receives ($15\%$ by default in the EBM implementation, as detailed in
Table~\ref{tab:EBM_parameters}) and stops boosting once the validation loss ceases to improve. It
is controlled by \texttt{early\_stopping\_rounds} and
\texttt{early\_stopping\_tolerance}. This internal validation partition is
always drawn from the data passed to the estimator, i.e. from the training
folds during 3-fold cross-validation, and from
$\mathcal{D}^{\text{train}}_{g}$ during the final re-estimation used to
assess performance on the test set. It is therefore disjoint from both the
validation fold used to select hyperparameters and the test set
$\mathcal{D}^{\text{test}}_{g}$. Consequently, the number of rounds is
selected without any access to $\mathcal{D}^{\text{test}}_{g}$.

\begin{table}[ht]
    \caption{Grid search and optimal hyperparameters for the claim severity and frequency EBM models.}
    \fontsize{8}{10}\selectfont
    \centering
    \renewcommand{\arraystretch}{1.3} 
    \setlength{\tabcolsep}{8pt}
    \begin{tabular} {lccccc}
    \toprule
        \textbf{Hyperparameter} & \textbf{Grid Severity} & \textbf{Optimal Severity} & \textbf{Grid Frequency} & \textbf{Optimal Frequency} \\ \midrule
        \texttt{smoothing\_rounds} & $[100, 200, 500]$ & 500 & $[100, 200, 500]$ & 200 \\ 
        \texttt{interactions} & $[0.5, 0.75, 0.9]$ & 0.9 & $[0.5, 0.75, 0.9]$ & 0.5 \\ 
        \texttt{max\_bins} & $[32, 1024, 4096]$ & 32 & $[32, 1024, 4096]$ & 32 \\ 
        \texttt{max\_interaction\_bins} & $[8, 16, 32]$ & 32 & 32 & 32 \\ 
        \texttt{outer\_bags} & $[14, 50]$ & 14 & 14 & 14 \\ 
        \texttt{learning\_rate} & $[0.005, 0.01, 0.02]$ & 0.02 & $[0.005, 0.01, 0.02]$ & 0.005 \\ 
        \texttt{max\_leaves} & $[2, 3, 4]$ & 3 & $[2, 3]$ & 3 \\ \bottomrule
    \end{tabular}
    \label{tab:grid_ebm}
\end{table}

\subsubsection{XGB}
\label{subsubsec:app_xgb}

The XGB model is implemented using the \textbf{xgboost} \texttt{Python} package \citep{chen_2016_xgboost}, with Poisson deviance as the loss function for the claim frequency model and gamma deviance for the claim severity model. To achieve optimal performance while mitigating overfitting, XGB hyperparameters are tuned via grid search, as described in Table~\ref{tab:grid_xgb}. In particular, the grid for the claim frequency model is designed to reduce the overfitting observed when using the default hyperparameter settings. 
We note that no early stopping is used in this procedure.

\begin{table}[ht]
    \caption{Grid search and optimal hyperparameters for the claim severity and frequency XGB models.}
    \centering    
    \fontsize{8}{10}\selectfont
    \renewcommand{\arraystretch}{1.3} 
    \setlength{\tabcolsep}{8pt}
    \begin{tabular} {lccccc}
    \toprule
        \textbf{Hyperparameter} & \textbf{Grid Severity} & \textbf{Optimal Severity} & \textbf{Grid Frequency} & \textbf{Optimal Frequency} \\ \midrule
         \texttt{n\_estimators} & $[100, 200, 300]$ & 200 & $[75, 100]$ & 100 \\ 
        \texttt{learning\_rate} & $[0.1, 0.3, 0.5]$ & 0.1 & 0.3 & 0.3 \\ 
         \texttt{max\_depth} & $[4, 6, 8]$ & 4 & $[2, 3]$ & 2 \\ 
        \texttt{min\_child\_weight} & $[1, 3]$ & 1 & $[10, 50, 100]$ & 100 \\ 
        \texttt{gamma} & $[0, 0.05]$ & 0 & 0 & 0 \\ 
        \texttt{alpha} & $[0, 1]$ & 1 & $[0, 5, 10]$ & 0 \\ 
        \texttt{lambda} & 0 & 0 & $[0, 5, 10]$ & 0 \\ 
        \texttt{subsample} & $[0.75, 1]$ & 1 & $[0.75, 1]$ & 0.75 \\ 
        \texttt{colsample\_bytree} & $[0.75, 1]$ & 0.75 & $[0.75, 1]$ & 1 \\ \bottomrule
    \end{tabular}
    \label{tab:grid_xgb}
\end{table}

\subsubsection{NAM}
\label{subsubsec:app_nam}

For the implementation of the NAM model \citep{agarwal_neural_2021}, we use the \textbf{NAMpy} \texttt{Python} package with Poisson deviance as the loss function for the claim frequency model and gamma deviance for the claim severity model. Consistent with standard practices in actuarial applications of neural networks, we employ the hyperbolic tangent activation function for the hidden layers and the exponential activation function for the output layer to implement a log-link. Given the high computational cost of the NAM model, performing a full grid search for hyperparameter tuning is impractical. Consequently, we adopt Bayesian optimization with sequential management to determine the optimal network configuration for fitting each shape function corresponding to each feature. Specifically, we perform 20 optimization iterations on a network trained for 500 epochs with a batch size of 8192, tuning the hyperparameters \texttt{hidden\_dims}, \texttt{feature\_dropout}, and \texttt{learning\_rate}, as summarized in Table~\ref{tab:bo_nam}. The hyperparameter \texttt{hidden\_dims} controls the number of neurons in the hidden layers \texttt{n\_layers} defining each feature network. The hyperparameter \texttt{feature\_dropout} specifies the proportion of feature networks randomly deactivated during training in order to reduce overfitting. 
\texttt{learning\_rate} determines the step size of the gradient-based optimization procedure. Finally, we adopt a fixed early-stopping criterion
that determines the number of epochs with no improvement before training is
stopped for the NAM model. Specifically, training is halted after 
20 iterations without an improvement greater than $10^{-4}$ in the loss.
This parameter is not subject to optimization.

\begin{table}[ht]
    \caption{Bayesian optimization for the claim severity and frequency NAM models.}
    \centering    
    \fontsize{8}{10}\selectfont
    \renewcommand{\arraystretch}{1.3} 
    \setlength{\tabcolsep}{8pt}
    \begin{tabular} {lccccc}
    \toprule
        \textbf{Hyperparameter} & \textbf{Min} & \textbf{Max} & \textbf{Optimal Severity} & \textbf{Optimal Frequency} \\ \midrule
        \texttt{n\_units} for \texttt{hidden\_dims} & 20 & 80 & 64 & 70 \\ 
        \texttt{n\_layers} for \texttt{hidden\_dims} & 2 & 5 & 2 & 3 \\ 
        \texttt{learning\_rate} & $10^{-5}$ & 0.01 & 0.01 & 0.0018 \\ 
        \texttt{feature\_dropout} & 0 & 0.01 & 0 & 0.0018 \\ \bottomrule
    \end{tabular}
    \label{tab:bo_nam}
\end{table}

\subsubsection{LocalGLMnet}
\label{subsubsec:app_lglm}

The LocalGLMnet model \citep{richman_LocalGLMnet_2023} is implemented in \textbf{TensorFlow} using \texttt{Python}, with Poisson deviance as the loss function for the claim frequency model and gamma deviance as the loss function for the claim severity model. The network weights are initialized from the coefficients of GLM models. Similarly to the NAM model described in Section~\ref{subsubsec:app_nam}, Bayesian optimization is employed to determine the neural network architecture, as detailed in Table~\ref{tab:bo_lglm}. The hyperbolic tangent activation function is used for the hidden layers, and the exponential activation function is applied at the output layer to implement a log-link. The optimized hyperparameters are \texttt{num\_neurons}, which controls the number of neurons in each hidden layer, \texttt{num\_dense\_layers}, which determines the number of hidden layers, \texttt{learning\_rate}, which governs the gradient descent step size during training, and \texttt{dropout\_rate}, which specifies the proportion of neurons randomly deactivated during training to reduce overfitting.

\begin{table}[ht]
    \caption{Bayesian optimization for the claim severity and frequency LocalGLMnet models.}
    \centering    
    \fontsize{8}{10}\selectfont
    \renewcommand{\arraystretch}{1.3} 
    \setlength{\tabcolsep}{8pt}
    \begin{tabular} {lccccc}
    \toprule
        \textbf{Hyperparameter} & \textbf{Min} & \textbf{Max} & \textbf{Optimal Severity} & \textbf{Optimal Frequency} \\ \midrule
        \texttt{num\_neurons} & 10 & 100 & 100 & 100 \\ 
        \texttt{num\_dense\_layers} & 4 & 10 & 5 & 8 \\ 
        \texttt{learning\_rate} & $10^{-5}$ & 0.01 & 0.001 & 0.0005 \\ 
        \texttt{dropout\_rate} & 0 & 0.1 & 0.08 & 0 \\ \bottomrule
    \end{tabular}
    \label{tab:bo_lglm}
\end{table}

\section{Additional results for claim frequency models}
\label{sec:app_results_freq}

This appendix reports the results obtained for the claim frequency models.
Performance metrics for all competing models are presented in
Table~\ref{tab:res_frequency}. 
In-sample and out-of-sample predictive
performance is evaluated in terms of both the RMSE and the Poisson deviance
loss. We note that any deviance loss is consistent for mean estimation and
can therefore be used for model validation. Overall, GAMboost, GAMboostLSS
and EBM with interactions appear to be the best-performing models in terms of
RMSE, although the predictive performance of all models remains very
similar. In terms of Poisson deviance, GAMboost and GAMboostLSS slightly outperform the other models.

\begin{table}[ht!]
\centering
\caption{
In-sample and out-of-sample prediction performance of the claim frequency models in terms of Poisson deviance and RMSE. For each model, the percentage column gives the relative gain (in \textcolor{teal}{green}) or loss (in \textcolor{red}{red}) over the reference GLMbins model.  \label{tab:res_frequency}
}
\fontsize{8}{10}\selectfont
\renewcommand{\arraystretch}{1.2} 
\setlength{\tabcolsep}{6pt}
\begin{tabular}{lcccccccccc}
\toprule
\textbf{Model} & \multicolumn{3}{c}{\textbf{Deviance}} & \multicolumn{3}{c}{\textbf{RMSE}} \\
 \cmidrule(lr){2-4} \cmidrule(lr){5-7} 
 & Train & Test & \% & Train & Test & \% \\
\midrule
GLMbins & 0.3312 & 0.3339 & 0 & 0.13827 & 0.13884 & 0 \\
GAM & 0.3315 & 0.3337 & \textcolor{teal}{-0.1\%} & 0.13827 & 0.13884 & \textcolor{orange}{0.00\%} \\ 
GAMboost & 0.3289 & \textbf{0.3303} & \textcolor{teal}{-1.1\%} & 0.13830 & \textbf{0.13883} & \textcolor{teal}{-0.01\%} \\
GAMboostLSS & 0.3289 & 0.3303 & \textcolor{teal}{-1.1\%} & 0.13830 & 0.13883 & \textcolor{teal}{-0.01\%} \\
cyc-GBM & 0.3291 & 0.3342 & \textcolor{red}{+0.1\%} & 0.13817 & 0.13886 & \textcolor{red}{+0.01\%} \\
EBM & 0.3311 & 0.3337 & \textcolor{teal}{-0.1\%} & 0.13826 & 0.13884 & \textcolor{orange}{0.00\%} \\
EBM$^2$ & 0.3307 & 0.3336 & \textcolor{teal}{-0.1\%} & 0.13824 & 0.13883 & \textcolor{teal}{-0.01\%} \\
XGB & 0.3305 & 0.3338 & \textcolor{orange}{0.0\%} & 0.13824 & 0.13884 & \textcolor{orange}{0.00\%} \\
NAM & 0.3325 & 0.3349 & \textcolor{red}{+0.3\%} & 0.13832 & 0.13887 & \textcolor{red}{+0.02\%} \\
LocalGLMnet & 0.3315 & 0.3345 & \textcolor{red}{+0.2\%} & 0.13827 & 0.13885 & \textcolor{red}{+0.01\%} \\ 
\bottomrule
\end{tabular}
\end{table}

The out-of-sample results of the Diebold--Mariano tests based on the RMSE are reported in Table~\ref{tab:DM_frequency}. For the claim frequency models, this metric does not clearly identify a single model as significantly more accurate than the others. However, the results indicate that the NAM model is significantly outperformed by all competing models, with the exception of the cyc-GBM model. 
Similar results are obtained using the Poisson deviance loss.

\begin{table}[ht!]
\fontsize{8}{10}\selectfont
\centering
\caption{Results of the Diebold-Mariano tests for the claim frequency models on the testing dataset
 in terms of RMSE.
For each pair consisting of a row model $F_A$ and a column model $F_B$, the null hypothesis $H_0$ is rejected if the corresponding $p$-value is below $0.05$ (highlighted in \textcolor{teal}{green}). In this case, model $F_B$ is considered statistically more accurate than model $F_A$.
 \label{tab:DM_frequency}}
\renewcommand{\arraystretch}{1.3} 
\setlength{\tabcolsep}{6pt}
\begin{tabular} {lcccccccccc}
\toprule
\textbf{Model} & GLMbins & GAM & GAMboost & GAMboostLSS & cyc-GBM & EBM & EBM$^2$ & XGB & NAM & LocalGLMnet \\ \midrule
GLMbins & & 0.185 & 0.233 & 0.253 & 0.864 & 0.438 & 0.129 & 0.539 & 0.993 & 0.745 \\
GAM & 0.815 & & 0.335 & 0.360 & 0.946 & 0.864 & 0.322 & 0.817 & 1.000 & 0.915 \\
GAMboost & 0.767 & 0.665 & & 1.000 & 0.914 & 0.783 & 0.588 & 0.785 & 1.000 & 0.906 \\
GAMboostLSS & 0.747 & 0.640 & \textcolor{teal}{0.000} & & 0.903 & 0.760 & 0.563 & 0.765 & 1.000 & 0.893 \\
cyc-GBM & 0.136 & 0.054 & 0.086 & 0.097 & & 0.102 & \textcolor{teal}{0.032} & 0.079 & 0.882 & 0.337 \\
EBM & 0.562 & 0.136 & 0.217 & 0.240 & 0.898 & & 0.093 & 0.594 & 0.999 & 0.801 \\
EBM$^2$ & 0.871 & 0.678 & 0.412 & 0.437 & 0.968 & 0.907 & & 0.878 & 1.000 & 0.932 \\
XGB & 0.461 & 0.183 & 0.215 & 0.235 & 0.921 & 0.406 & 0.122 & & 0.993 & 0.783 \\
NAM & \textcolor{teal}{0.007} & \textcolor{teal}{0.000} & \textcolor{teal}{0.000} & \textcolor{teal}{0.000} & 0.118 & \textcolor{teal}{0.001} & \textcolor{teal}{0.000} & \textcolor{teal}{0.007} & & \textcolor{teal}{0.006} \\
LocalGLMnet & 0.255 & 0.085 & 0.094 & 0.107 & 0.663 & 0.199 & 0.068 & 0.217 & 0.994 & \\ \bottomrule
\end{tabular}
\end{table}

\begin{figure}[ht!]
\centering
\begin{subfigure}{.33\textwidth}
  \centering
  \includegraphics[width=\linewidth]{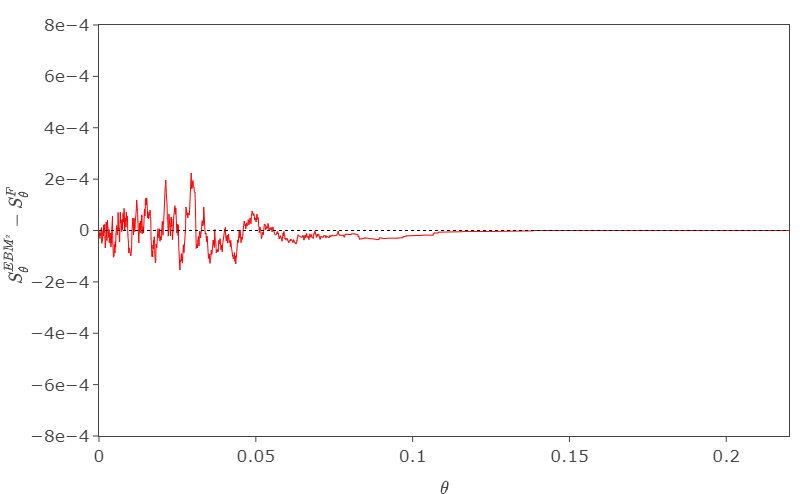}
  \caption{GLMbins}
  \label{fig:murphy_freq_GLM}
\end{subfigure}%
\begin{subfigure}{.33\textwidth}
  \centering
  \includegraphics[width=\linewidth]{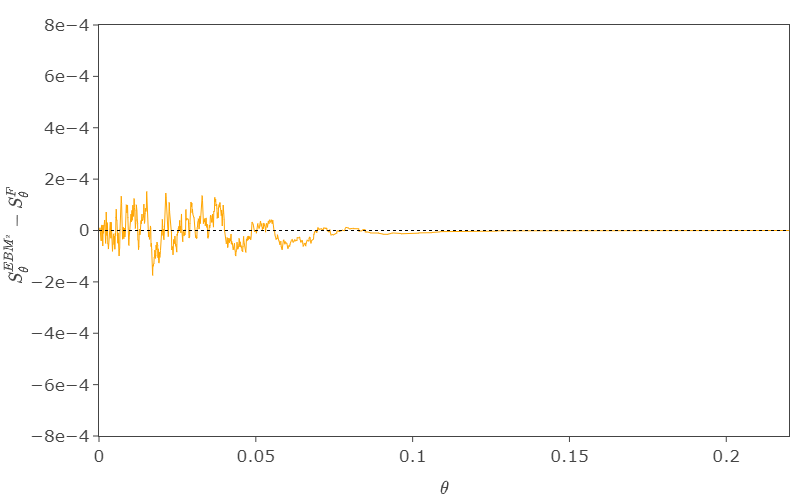}
  \caption{GAM}
  \label{fig:murphy_freq_GAM}
\end{subfigure}%
\begin{subfigure}{.33\textwidth}
  \centering
  \includegraphics[width=\linewidth]{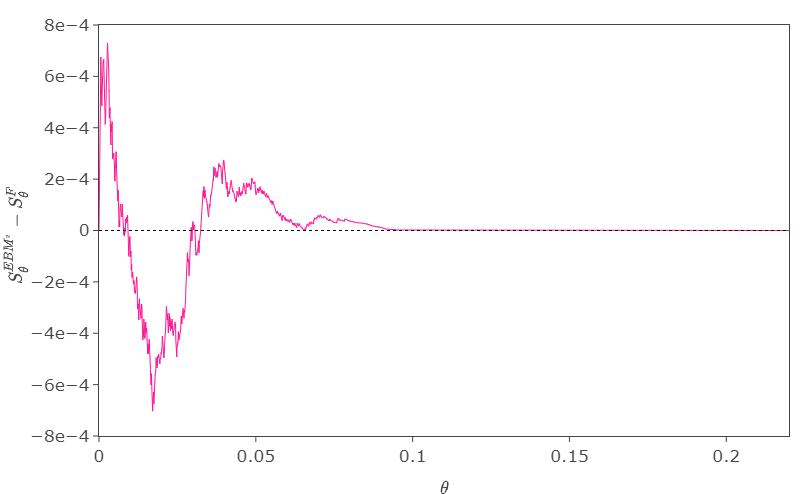}
  \caption{GAMboost}
  \label{fig:murphy_freq_GAM_Boost}
\end{subfigure}
\begin{subfigure}{.33\textwidth}
  \centering
  \includegraphics[width=\linewidth]{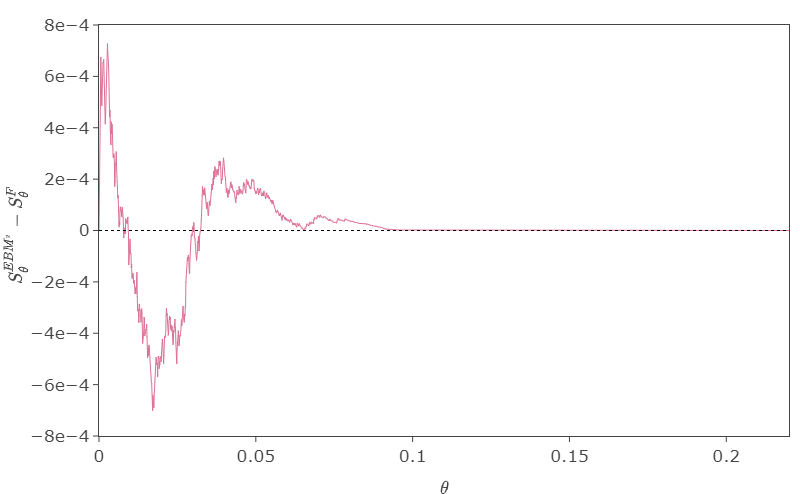}
  \caption{GAMboostLSS}
  \label{fig:murphy_freq_GAM_Boost_LSS}
\end{subfigure}%
\begin{subfigure}{.33\textwidth}
  \centering
  \includegraphics[width=\linewidth]{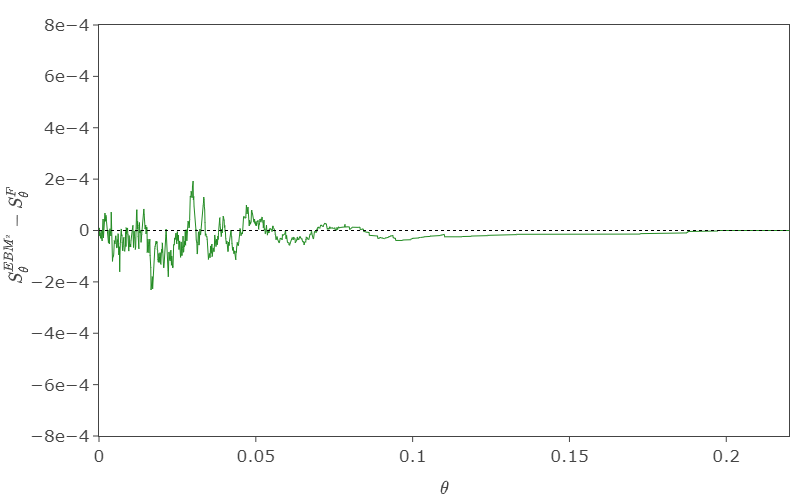}
  \caption{cyc-GBM}
  \label{fig:murphy_freq_GBM}
\end{subfigure}%
\begin{subfigure}{.33\textwidth}
  \centering
  \includegraphics[width=\linewidth]{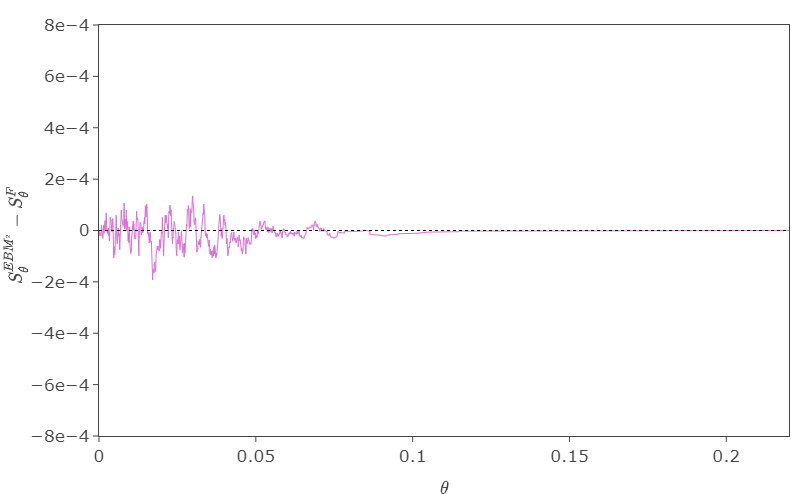}
  \caption{EBM}
  \label{fig:murphy_freq_EBM}
\end{subfigure}
\begin{subfigure}{.33\textwidth}
  \centering
  \includegraphics[width=\linewidth]{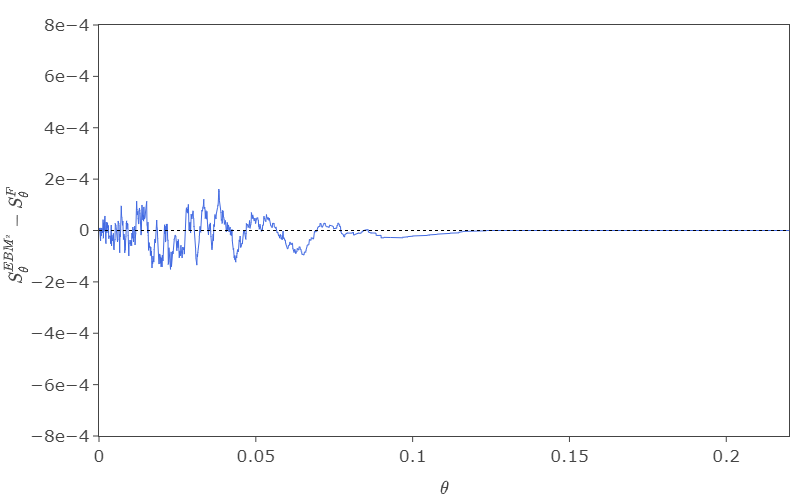}
  \caption{XGB}
  \label{fig:murphy_freq_XGB}
\end{subfigure}%
\begin{subfigure}{.33\textwidth}
  \centering
  \includegraphics[width=\linewidth]{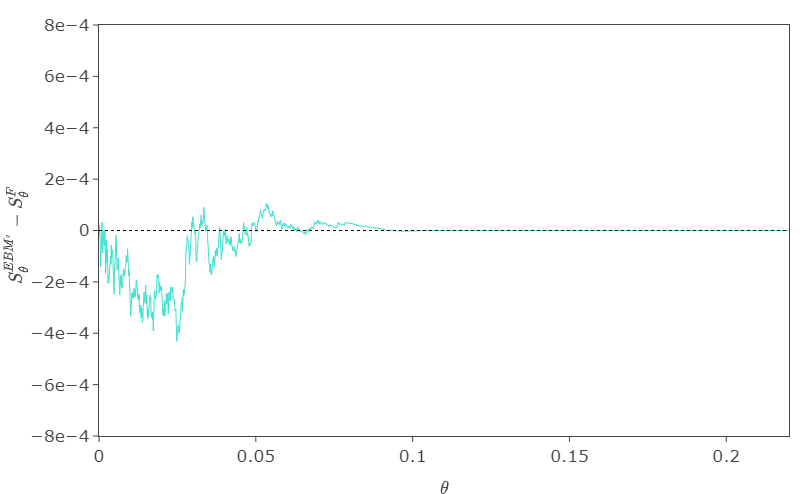}
  \caption{NAM}
  \label{fig:murphy_freq_NAM}
\end{subfigure}
\begin{subfigure}{.33\textwidth}
  \centering
  \includegraphics[width=\linewidth]{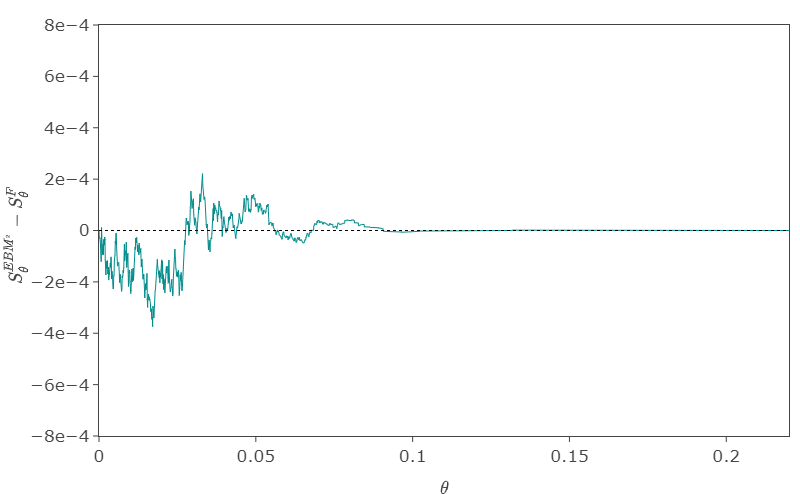}
  \caption{LocalGLMNet}
  \label{fig:murphy_freq_LGLM}
\end{subfigure}
\caption{
Murphy diagrams for the claim frequency models on the test dataset,
based on the difference between the elementary scoring function $S_{\theta}$
of the EBM$^2$ model and that of each competing model across the range of
parameter values $\theta$. A positive difference indicates that the competing
model outperforms EBM$^2$, whereas a negative difference indicates that
EBM$^2$ outperforms the competing model.
}
\label{fig:murphy_freq}
\end{figure}

To avoid relying on a specific loss function, the predictive dominance of
the competing claim frequency models is further examined using Murphy
diagrams, displayed in Figure~\ref{fig:murphy_freq}. As for claim severity models,
these diagrams report the difference between the elementary scoring function
$S_{\theta}$ of EBM$^2$ and that of each competing model across the range of
parameter values $\theta$, evaluated on the test dataset. We observe that
the curves of all models oscillate around $0$, indicating the absence of
uniform dominance over EBM$^2$, which is consistent with the previous
results. An interesting pattern emerges for the GAMboost and GAMboostLSS
models in Figures~\ref{fig:murphy_freq_GAM_Boost}
and~\ref{fig:murphy_freq_GAM_Boost_LSS}, respectively. These models are
outperformed by EBM$^2$ for values of $\theta$ approximately between $0.01$
and $0.03$, but dominate it for lower and higher values of $\theta$.

Finally, the results of the Bregman dominance test are reported in Table~\ref{tab:BD_frequency}. The GAMboost and GAMboostLSS models stand out in the comparison. Because these models can appear both significantly dominant and dominated depending on the region of the loss function considered, the null hypothesis of the Bregman dominance test can be rejected at the $5\%$ significance level whether they play the role of model $F_A$ or $F_B$. Aside from these two models, NAM and LocalGLMnet appear to be the least competitive, while it remains difficult to clearly identify a single model as systematically more predictive than the others. It is also worth noting the competitive performance of GLMbins, which may be explained by its initialization based on the EBM model.

\begin{table}[ht!]
\fontsize{8}{10}\selectfont
\centering
\caption{
Results of the Bregman dominance tests for the claim frequency models on the testing dataset. For each pair consisting of a row model $F_A$ and a column model $F_B$, the null hypothesis $H_0$ is rejected if the corresponding $p$-value is below $0.05$ (highlighted in \textcolor{teal}{green}). In this case, model $F_B$ is considered statistically more accurate than model $F_A$.  \label{tab:BD_frequency}}
\renewcommand{\arraystretch}{1.3} 
\setlength{\tabcolsep}{6pt}
\begin{tabular} {lcccccccccc}
\toprule
\textbf{Model} & GLMbins & GAM & GAMboost & GAMboostLSS & cyc-GBM & EBM & EBM$^2$ & XGB & NAM & LocalGLMnet \\ \midrule
GLMbins & & 0.242 & \textcolor{teal}{0.000} & \textcolor{teal}{0.000} & 0.816 & 0.860 & 0.864 & 0.212 & 1.000 & 0.682 \\
GAM & 0.804 & & \textcolor{teal}{0.000} & \textcolor{teal}{0.000} & 0.832 & 0.950 & 0.658 & 0.958 & 1.000 & 0.668 \\
GAMboost & \textcolor{teal}{0.000} & \textcolor{teal}{0.000} & & 0.928 & \textcolor{teal}{0.002} & \textcolor{teal}{0.000} & \textcolor{teal}{0.000} & \textcolor{teal}{0.000} & \textcolor{teal}{0.016} & \textcolor{teal}{0.008} \\
GAMboostLSS & \textcolor{teal}{0.000} & \textcolor{teal}{0.000} & 0.220 & & \textcolor{teal}{0.000} & \textcolor{teal}{0.000} & \textcolor{teal}{0.000} & \textcolor{teal}{0.000} & \textcolor{teal}{0.024} & \textcolor{teal}{0.010} \\
cyc-GBM & 0.182 & 0.408 & \textcolor{teal}{0.000} & \textcolor{teal}{0.000} & & 0.162 & 0.098 & 0.636 & 0.992 & 0.812 \\
EBM & 0.318 & \textcolor{teal}{0.030} & \textcolor{teal}{0.000} & \textcolor{teal}{0.000} & 0.628 & & 0.068 & 0.246 & 0.990 & 0.554 \\
EBM$^2$ & 0.158 & 0.854 & \textcolor{teal}{0.000} & \textcolor{teal}{0.000} & 0.362 & 0.656 & & 0.636 & 0.992 & 0.598 \\
XGB & 0.070 & 0.214 & \textcolor{teal}{0.000} & \textcolor{teal}{0.000} & 0.784 & 0.084 & 0.722 & & 0.994 & 0.852 \\
NAM & \textcolor{teal}{0.002} & \textcolor{teal}{0.000} & \textcolor{teal}{0.000} & \textcolor{teal}{0.000} & \textcolor{teal}{0.002} & \textcolor{teal}{0.002} & \textcolor{teal}{0.000} & \textcolor{teal}{0.000} & & \textcolor{teal}{0.024} \\
LocalGLMnet & \textcolor{teal}{0.002} & \textcolor{teal}{0.006} & \textcolor{teal}{0.000} & \textcolor{teal}{0.000} & \textcolor{teal}{0.048} & \textcolor{teal}{0.006} & \textcolor{teal}{0.002} & \textcolor{teal}{0.010} & 0.992 & \\ \bottomrule
\end{tabular}
\end{table}

\begin{figure}[ht!]
    \centering
    \includegraphics[width=0.7\linewidth]{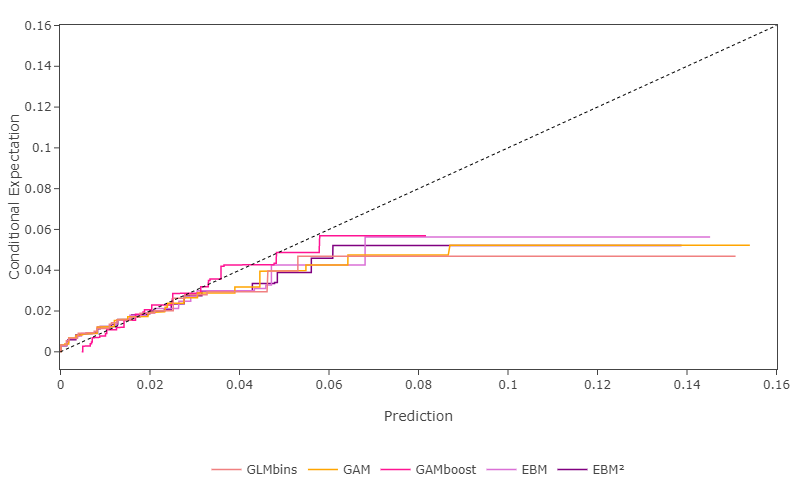}
    \caption{Calibration plot for the claim frequency GLMbins, GAM, GAMboost, EBM and EBM$^2$ models on the testing dataset.}
    \label{fig:isotonic_freq}
\end{figure}

Regarding the calibration assessment, T-reliability diagrams are presented in Figure \ref{fig:isotonic_freq} for GLMbins, GAM, GAMboost, and EBM, both with and without interactions. 
The models are well calibrated up to approximately $0.03$. Beyond this
point, the calibration begins to deteriorate gradually, with the departures
becoming pronounced from around $0.06$ onwards. These deviations arise in
the right tail of the distribution, where observations are sparse and the
calibration curve is unstable. Indeed, this region accounts for only
a small fraction of the test sample.
According to this graphical analysis, the GAMboost model exhibits the best calibration.  
Consequently, the null hypothesis of the auto-calibration test is rejected for all competing models at the $0.5\%$ significance level, including GLMbins.
Finally, the Murphy decomposition reported in Table~\ref{tab:murphy_dec_freq} identifies the GAMboost model as the least miscalibrated and the best overall when jointly accounting for miscalibration, discrimination, and uncertainty, closely followed by GAMboostLSS and EBM with interactions. GLMbins and EBM with interactions show the strongest discrimination abilities.

\begin{table}[ht!]
    \caption{Murphy decomposition for the claim frequency models on the testing dataset.}
    \fontsize{8}{10}\selectfont
    \centering
    \renewcommand{\arraystretch}{1.3} 
    \setlength{\tabcolsep}{6pt}
    \begin{tabular} {lccccc}
    \toprule
    \textbf{Model} & \textbf{Miscalibration} & \textbf{Discrimination} & \textbf{Uncertainty} & \textbf{Score} \\ \midrule
    GLMbins & 9.7 $\times$ 10$^{-5}$ & \textbf{8.0 $\times$ 10$^{-5}$} & 1.9260 $\times$ 10$^{-2}$ & 1.9277 $\times$ 10$^{-2}$ \\ 
    GAM & 9.4 $\times$ 10$^{-5}$ & 7.8 $\times$ 10$^{-5}$ & 1.9260 $\times$ 10$^{-2}$ & 1.9275 $\times$ 10$^{-2}$ \\ 
    GAMboost & \textbf{8.6 $\times$ 10$^{-5}$} & 7.2 $\times$ 10$^{-5}$ & 1.9260 $\times$ 10$^{-2}$ & \textbf{1.9273 $\times$ 10$^{-2}$} \\ 
    GAMboostLSS & 8.6 $\times$ 10$^{-5}$ & 7.2 $\times$ 10$^{-5}$ & 1.9260 $\times$ 10$^{-2}$ & 1.9273 $\times$ 10$^{-2}$ \\ 
    cyc-GBM & 9.9 $\times$ 10$^{-5}$ & 7.7 $\times$ 10$^{-5}$ & 1.9260 $\times$ 10$^{-2}$ & 1.9281 $\times$ 10$^{-2}$ \\ 
    EBM & 9.4 $\times$ 10$^{-5}$ & 7.7 $\times$ 10$^{-5}$ & 1.9260 $\times$ 10$^{-2}$ & 1.9277 $\times$ 10$^{-2}$ \\ 
    EBM$^2$ & 9.3 $\times$ 10$^{-5}$ & 7.9 $\times$ 10$^{-5}$ & 1.9260 $\times$ 10$^{-2}$ & 1.9274 $\times$ 10$^{-2}$ \\ 
    XGB & 9.5 $\times$ 10$^{-5}$ & 7.8 $\times$ 10$^{-5}$ & 1.9260 $\times$ 10$^{-2}$ & 1.9277 $\times$ 10$^{-2}$ \\ 
    NAM & 9.7 $\times$ 10$^{-5}$ & 7.0 $\times$ 10$^{-5}$ & 1.9260 $\times$ 10$^{-2}$ & 1.9286 $\times$ 10$^{-2}$ \\ 
    LocalGLMnet & 9.7 $\times$ 10$^{-5}$ & 7.7 $\times$ 10$^{-5}$ & 1.9260 $\times$ 10$^{-2}$ & 1.9279 $\times$ 10$^{-2}$ \\ \bottomrule
    \end{tabular}
    \label{tab:murphy_dec_freq}
\end{table}

\begin{figure}[ht!]
    \centering
    \begin{subfigure}{.33\textwidth}
    	\centering
    	\includegraphics[width=\linewidth]{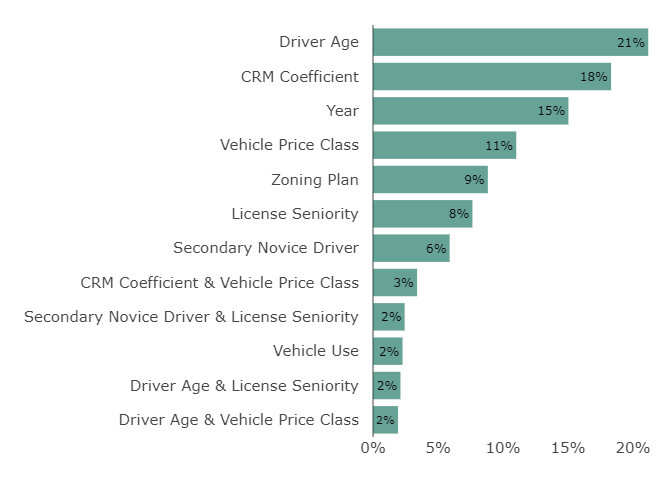}
    	\caption{Relative feature importance $RI_j$}
    	\label{fig:global_EBM_freq}
    \end{subfigure}%
	\begin{subfigure}{.33\textwidth}
  		\centering
  		\includegraphics[width=\linewidth]{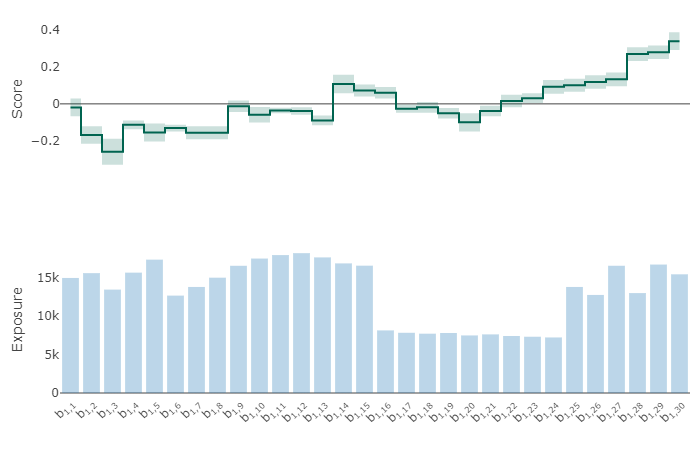}
  		\caption{Driver Age}
  		\label{fig:shape_age_COND_freq}
  	\end{subfigure}%
 	\begin{subfigure}{.33\textwidth}
 		\centering
 		\includegraphics[width=\linewidth]{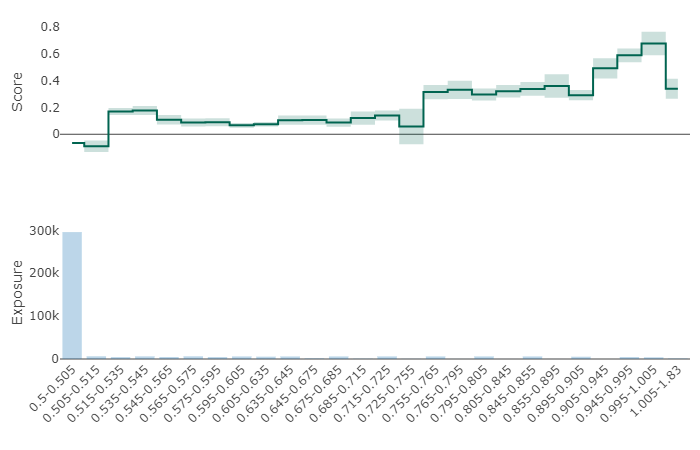}
 		\caption{CRM Coefficient}
 		\label{fig:shape_cof_CRM_freq}
 	\end{subfigure}   
\caption{
Relative feature importance in the EBM$^2$ claim frequency model, with features sorted by descending importance (Figure~\ref{fig:global_EBM_freq}), and predicted shape functions $f_j(\mathbf{x}_i)$ for Driver Age (Figure~\ref{fig:shape_age_COND_freq}) and CRM Coefficient (Figure~\ref{fig:shape_cof_CRM_freq}). Shaded areas indicate uncertainty bands estimated via bagging.
}
\label{fig:shape_freq}
\end{figure}

\begin{figure}[ht!]
\centering
\begin{subfigure}{.5\textwidth}
  \centering
  \includegraphics[width=\linewidth]{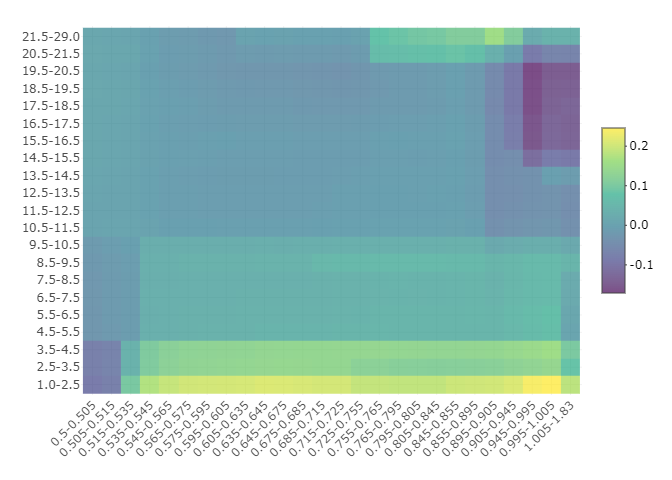}
  \caption{CRM Coef. $\times$ Veh. Price Class}
  \label{fig:heatmap_cof_CRM_prix_SRA_freq}
\end{subfigure}%
\begin{subfigure}{.5\textwidth}
  \centering
  \includegraphics[width=\linewidth]{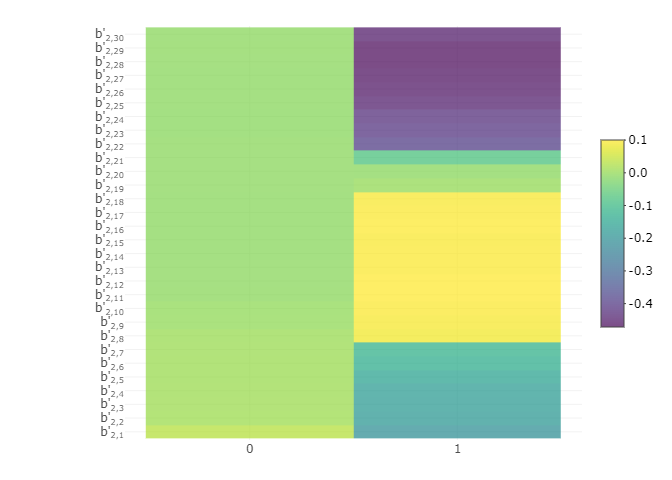}
  \caption{2nd Nov. Driver $\times$ License Senior.}
  \label{fig:heatmap_novice_COND2_anc_COND_freq}
\end{subfigure}
\caption{Heatmaps of the predicted pairwise interaction terms $f_{(j_1,j_2)}$ in the claim frequency EBM$^2$ model. The displayed interactions correspond to (CRM Coefficient, Vehicle Price Class) in Figure~\ref{fig:heatmap_cof_CRM_prix_SRA_freq} and (Second Novice Driver, License Seniority) in Figure~\ref{fig:heatmap_novice_COND2_anc_COND_freq}. Color intensity reflects the magnitude of the interaction effect on predicted claim frequency.}
\label{fig:heatmap_freq}
\end{figure}

To conclude this appendix, we present the interpretability results of the EBM$^2$ claim frequency model. Global effects on the training dataset are illustrated in terms of variable importance in Figure \ref{fig:global_EBM_freq}, while local effects are shown through examples of main effect terms for Driver Age and CRM Coefficient in Figures\ref{fig:shape_age_COND_freq} and~\ref{fig:shape_cof_CRM_freq}, respectively. Pairwise interactions for (CRM Coefficient, Vehicle Price Class) and (Secondary Novice Driver, License Seniority) are displayed in Figures~\ref{fig:heatmap_cof_CRM_prix_SRA_freq} and~\ref{fig:heatmap_novice_COND2_anc_COND_freq}.

